\documentclass[12pt,a4paper]{article}
\pdfoutput=1
 \usepackage{jheppub}
 \usepackage{amsmath}
 \usepackage{amsfonts}
 \usepackage{mathtools}
 \usepackage{amssymb}
 \usepackage{caption}
 \usepackage{subcaption}
 \usepackage{slashed}
\usepackage{float}
\usepackage{braket}
\usepackage{graphicx}
\usepackage{etoolbox}
\graphicspath{ {heterotic} }

\def\b{\bullet}
\def\al{\alpha}

\newcommand{\be} {\begin{equation}}
\newcommand{\ee} {\end{equation}}
\newcommand{\bea} {\begin{eqnarray}}
\newcommand{\eea} {\end{eqnarray}}
\newcommand{\ba} {\begin{array}}
\newcommand{\ea} {\end{array}}
\newcommand{\nn} {\nonumber}

 \title{Gravitational couplings in ${\cal N}=2$ heterotic compactifications with Wilson lines}
 \author{Aradhita Chattopadhyaya}
\affiliation{School of Mathematics, Trinity College Dublin, \\ Dublin 2, Ireland}
\emailAdd{aradhita@maths.tcd.ie}

\abstract{In this paper we compute the gravitational couplings of the heterotic string compactified on $(K3\times T^2)/\mathbb{Z}_N$ and $E_8\times E_8$ and predict the Gopakumar Vafa invariants of the dual Calabi Yau manifold in presence of Wilson lines. Here $\mathbb{Z}_N$ acts as an automorphism  on $K3$ associated with the conjugacy classes of $M_{23}$ and a shift of $1/N$ on one of the $S^1$ of $T^2$. We study in detail the cases $N=2,3$ for standard and several non-standard embeddings where $K3$ is realized as toroidal orbifolds $T^4/\mathbb{Z}_4$ and $T^4/\mathbb{Z}_3$. From these computations we extract the polynomial term in perturbative pre-potential for these orbifold models in presence of a single Wilson line. We also show for standard embeddings the integrality of the Gopakumar Vafa invariants depend on the integrality of Fourier coefficients of Fourier transform of the twisted elliptic genus of $K3$ in presence of $n<8$ Wilson lines.}

\begin{document}
\maketitle
\flushbottom

\section{Introduction}

In the seminal work \cite{Harvey-Moore} the authors relate the elliptic genus of $K3$ under spectral flow symmetry to the new-supersymmetric index \cite{Cecotti:1992qh,Cecotti:1992vy}. Several symplectic automorphisms of $K3$ corresponding to Mathieu group $M_{24}$ was noted in the earlier works of \cite{Gaberdiel:2010ch,Gaberdiel:2011fg,Eguchi:2010ej,Eguchi:2010fg,Cheng:2010pq,Cheng:2013kpa,David:2006ji}. We generalized the relation between twisted elliptic genus of $K3$ and the new super-symmetric index in the previous work \cite{Chattopadhyaya:2016xpa} to different orbifolds of $K3$ while the embeddings were of standard type where one $SU(2)$ of one of the $E_8$ lattices was coupled to the bosons of $K3$. This new super-symmetric index is the most important ingredient in predicting the Gopakumar Vafa invariants corresponding to Calabi Yau geometry in its dual type II theory. Given the twisted elliptic genus we can evaluate these Gopakumar Vafa invariants of the dual Calabi Yau manifolds for the standard embeddings \cite{Chattopadhyaya:2017zul}. In the present work we show that the integrality of the Gopakumar Vafa invariants under standard embeddings depend solely on the integrality of the Fourier transform of the twisted elliptic genus of $K3$ for these orbifolds. This is true even when several Wilson lines $(n<8) $ are added. We show for all orbifolds of $K3$ where the orbifold action $g'$ corresponds to the conjugacy classes of $M_{24}$. It is interesting to observe the fact that the Fourier transform of the twisted elliptic genus of $K3$ should integer coefficients is also demanded by their corresponding Siegel modular forms (obtained as theta lift of the twisted elliptic genus) to have integer Fourier coefficients. These in turn would count the number of $1/4$ BPS states for ${\cal N}=4$ type IIB theory on $(K3\times T^2)/\mathbb{Z}_N$ \cite{Sen:2010mz,Chattopadhyaya:2017ews,Dijkgraaf:1996it}.

Through the computation of the purely holomorphic part of gravitational coupling $F_g$ which appears in the low energy effective action of the theory we can predict the genus $g$ Gopakumar Vafa invariants $n_{m}^g$ from $F_g^{\rm GV}$.
\bea \label{fgv}
F_g^{{\rm GV}} &=& \frac{(-1)^{g+1}}{2(2\pi )^{2g-2}} \bar F_g^{\rm hol},\\ \nn
F^{{\rm GV}}_g&=&\frac{(-1)^g |B_{2g}B_{2g-2}|\chi(X)}{4g (2g-2)(2g-2)!}\\ \nn
&+& \sum_{m}\left[\frac{|B_{2g}|n_m^0}{2g (2g-2)!}+\frac{2(-1)^g n_m^2}{(2g-2)!}\pm...-\frac{g-2}{12}n_{m}^{g-1}+n_m^g\right]{\rm Li}_{3-2g}(e^{2\pi i m\cdot y}).
\eea
It is interesting to generalize the canonical well-studied ${\cal N}=2$ duality of heterotic on $K3\times T^2$ and type IIA studied in \cite{Kachru:1995wm,Klemm:1995tj} to these orbifolds of $K3$. Due to presence of  unbroken gauge group $E_8$ it is computationally more involved to compute the results from the type II side under standard embeddings.
Non-standard embeddings therefore play an interesting role as a check for the heterotic type II duality symmetries. In the previous works \cite{Weiss:2007tk,Banlaki:2018pcc,Banlaki:2019bxr} some non-standard embeddings introduced in \cite{Stieberger:1998yi,Honecker:2006qz} were studied and their Gopakumar Vafa invariants prediction from the heterotic computation was matched by direct computation in the type II side. In this work we extend the heterotic prediction with the Wilson line addition to these orbifold models. For the non-standard embeddings we take the heterotic string to be compactified on $(T^4/\mathbb{Z}_\nu\times T^2)/\mathbb{Z}_N$. Here $\nu=4$ when $N=2$ and $\nu=3$ where $N=3$. We observe here that for single Wilson line being present the conifold singularities where the vectors become massless originate only in the sector where the charges associated to the Wilson line moduli are even. To obtain these results we used the generalized hatting procedure which illustrates how the $E_8$ lattice is broken by the addition of Wilson lines.
% We have used the generalized hatting procedure as in \cite{Weiss:2007tk} and the conifold singularities are obtained from the $E_{4,1}^{even}$ part of the new supersymmetric index.
In general for $n$ Wilson lines the classical vector multiplet moduli space is given by the Kahler space of 
\[\frac{SU(1,1)}{U(1)}\times \frac{SO(2+n,2)}{SO(2+n)\times SO(2)}. \]
The first factor corresponds to dilaton and the second factor gives the Wilson line and torus moduli. The vector multiplet couplings transform as automorphic functions under subgroups of $SO(2+n,2,\mathbb{Z})$.

Another important aspect of the heterotic type II duality symmetry involves the matching of the perturbative pre-potential on both sides of the duality. In the previous works of \cite{Harvey-Moore,LopesCardoso:1995nk,LopesCardoso:1996nc} the perturbative pre-potential of the ${\cal N}=2$ heterotic strings compactified on $K3\times T^2$ and two $E_8$ lattices were computed for different instanton embeddings. This pre-potential contains a polynomial term which can be computed from the heterotic one loop gravitational coupling corrections. This involves computing the $F_g$ in degenerate orbit in absence of any gravi-photon field strength. In \cite{Harvey-Moore} this quantity was calculated in presence of 0 or 8 Wilson lines moduli, and in \cite{LopesCardoso:1995nk,LopesCardoso:1996nc} it was computed for a single Wilson line for heterotic string compactified on $K3\times T^2$ and $E_8\times E_8$.  The computation involved evaluation of the one-loop gravitational couplings using the unfolding technique similar to \cite{Dixon:1990pc}. In the present work we generalize these computations to orbifolds of $K3$ by $g'$ where $g'\in [M_{23}]$ and a shift of $1/N$ in one of the circles of $T^2$.
%In \cite{Harvey-Moore,LopesCardoso:1995nk,LopesCardoso:1996nc}  the authors discuss computation of the polynomial part of the perturbative pre-potential for heterotic compactified on $K3\times T^2$ 

For all the compactifications we study in this paper we can compute $F_g$ in the degenerate orbit at genus 1 using the method described in \cite{Harvey-Moore,Borcherds:1996uda,David:2006ji,Chattopadhyaya:2017zul}. This result can be written in manner from which one can extract the polynomial term as
\[-\frac{1}{2Y}d_{abc}y_2^ay_2^by_2^c,\]
where $y=\{T,U,\vec V\}$, $y_2={\rm Im}(y)$. $d_{abc}$ can be matched with the triple intersection numbers of the Calabi Yau geometry. These in general depend on instanton embeddings. $d_{abc}$ would in general also depend on the chamber of the moduli space where the gravitational threshold is computed. In this paper we shall only refer to the region $T_2/N>U_2$ and $|V_2|<<U_2$ in the STUV model. The main ingredient for this computation is again the new super-symmetric index.

\vspace{0.3cm}
\noindent
The organization of the paper is as follows:
In the next section \ref{stand} we introduce the standard embeddings and the generalized hatting procedure following \cite{Harvey-Moore,Weiss:2007tk,Chattopadhyaya:2016xpa} and relate the twisted elliptic genus and new supersymmetric index. We do this for a general number of Wilson lines $n\le 8$. In the earlier work \cite{Chattopadhyaya:2016xpa} this relation was obtained for single Wilson line. The new supersymmetric index is the main ingredient for computing the Gopakumar Vafa invariants of the dual Calabi Yau 3-folds.
 Then in section \ref{gvextract} we compute the Gopakumar Vafa invariants from the new supersymmetric index following the previous works of \cite{Harvey-Moore,Borcherds:1996uda,David:2006ji,Marino:1998pg,Marino:2002wa,Gopakumar:1998ii,Gopakumar:1998jq,Chattopadhyaya:2017zul}. The results in this section would hold for any number of Wilson lines $n<8$.
In section \ref{prepot} we compute the pre-potential polynomial for the heterotic string compactified on $\mathbb{Z}_N$ orbifolds of $K3\times T^2$ for the moduli $T,U,V$ from the heterotic side. It's observed to be rational multiple of $2\pi$ as expected from the predictions of \cite{Harvey-Moore}. We illustrate these results for order 2 and 3 orbifolds of $K3$ for various non-standard embeddings studied in \cite{Chattopadhyaya:2016xpa,Banlaki:2018pcc}.
In section \ref{con} we conclude with discussions on and comparison with existing literature and possible future directions.
In the appendix \ref{znew3} we briefly demonstrate the evaluation of new supersymmetric index for heterotic compactified on $K3\times T^2$ where $K3$ is realized as $T^4/\mathbb{Z}_3$ and $\mathbb{Z}_3$ orbifold acts on the orbifold of torus with a $1/3$ shift on one of the $T^2$ circles. Appendix \ref{gv} lists the genus zero Gopakumar-Vafa invariants for standard and several non-standard embeddings of $K3$ with orbifolds of order 2 and 3.
Appendix \ref{lowcexp} lists the low lying coefficients of the new super-symmetric index and also the first few terms of discrete Fourier transform of the twisted elliptic genus of $K3$ for order 2 and 3 orbifolds of $K3$.
%In this work we generalize the 

\section{Standard embeddings and New super-symmetric index}\label{stand}
In this section we argue that for all standard embeddings for various $\mathbb{Z}_N$ orbifolds of $K3$ where the orbifold action acts as a $g'$ action on $K3$ where $g'\in [M_{24}]$ the genus zero Gopakumar Vafa invariants are integral. This predicts that under the action of the Wilson lines dual Calabi Yau geometries exist. The integrality of these Gopakumar Vafa invariants are ensured by the integrality of the Fourier transform of the Fourier coefficients of twisted elliptic genus of $K3$.

\subsection{Definition of standard embeddings}
In heterotic strings compactified on $E_8\times E_8$ and $(K3\times T^2)/{\mathbb{Z}_N}$ the spin connection the $K3$ bosons is coupled to the gauge connection of one of the $SU(2)$s of one $E_8$ lattice. The $\mathbb{Z}_N$ action is an order $N$ action corresponding to 26 conjugacy classes of $M_{24}$ Mathieu group and a shift of $1/N$ in one of the $S^1$ of the $T^2$.

If the two $E_8$ lattices are realized by $16 \oplus 16$ fermions (in the absence of Wilson lines) then the gauge connection for standard embedding is given as:
\begin{equation}
{\cal G} =  \sum_{I, J  =1}^4 \lambda^{ I}  B_a^{ I  J} \partial X^a  \lambda^  J
+ \sum_{I, J = 5}^{16} \lambda^I A_i^{IJ} \partial X^i \lambda^J 
+ \sum_{I, J=1}^{16}
\lambda^{\prime I} A_{i} ^{\prime IJ} \partial X^i \lambda^{\prime J} ,
\end{equation}
where $A_i, A_i'$ are flat connections of the two torus $T^2$, $B_a$ is the $SU(2)$ spin connection on $T^4$, $\lambda^I, \, \lambda'^J$ are coordinates of the two $E_8$ fermions one of which is coupled to $K3$ via one $SU(2)$.
This embedding breaks the $E_8$ lattice to a $D2$ (or $SO(4)$) and $D6$ ($SO(12)$) while the other $E_8$ remains completely untouched. The 4 interacting Majorana-Weyl fermions from one $E_8$ coupled to the 4 bosons of the $K3$ produces a super-conformal field theory on $K3$. This $K3$ is orbifolded by the action of $g'$ corresponding to $[M_{24}]$.
In this paper we only deal with the cases where all the Wilson lines are present in the $E_8$ lattice which is not coupled to the $K3$.

\noindent
The internal CFT is split as follows:
\begin{equation} \label{decomp}
{\cal H}^{internal} = {\cal H}_{D2K3}
\otimes  {\cal H}_{D6} \otimes 
{\cal H}_{T^2}  \otimes [{\cal H} _{E_8}] .
\end{equation}
The last term would be modified depending on the presence of different number of Wilson lines in the picture.
For $n$ Wilson lines where $n<8$ it is given by,
\begin{equation} \label{decompwil}
{\cal H}^{internal} = {\cal H}_{D2K3}
\otimes  {\cal H}_{D6} \otimes 
{\cal H}_{T^2}  \otimes [{\cal H} ^{(8-n , 0)}_{E_{8,n}}] .
\end{equation}
where, $E_{8,n}$ is the broken $E_8$ lattice due to addition of Wilson lines. $g'$ is given by a $\mathbb{Z}_N$ automorphism on the $(6,6)$ CFT of ${\cal H}_{D2K3}$ and a $1/N$ shift acts on the $T^2$.

\subsection{New supersymmetric index and twisted elliptic genus}
Here we briefly review the relation between the new supersymmetric index \cite{Cecotti:1992qh} and twisted elliptic genus of $K3$ \cite{Eguchi:2010ej,Eguchi:2010fg,Cheng:2010pq,Cheng:2013kpa,Gaberdiel:2010ch,Gaberdiel:2011fg} following the arguments of \cite{Harvey-Moore,Chattopadhyaya:2016xpa}.
The new supersymmetric index is given by the following trace
\begin{equation} \label{znewtr}
{\cal Z}_{{\rm new}}=\frac{1}{\eta^2}{\rm Tr}_R((-1)^{F_R} F_R q^{L_0-c/24}\bar{q}^{\bar{L_0}-\bar{c}/24}),
\end{equation}
where $F_R$ is the total fermion number from the right moving sectors. However due to the zero modes of $T^2$ only $F^{T^2}_0$ contributes and this gives:
\begin{equation} \label{decompk3}
{\cal Z}_{\rm new} = 
\frac{1}{\eta^2(\tau) } 
\sum_{r,s=0}^{N-1}\left [  \frac{\theta_2^6(\tau)}{\eta^6(\tau)}  \Phi^{(r, s) }_R  +
\frac{  \theta_3^6(\tau)}{\eta^6(\tau) } \Phi^{(r, s) }_{NS^+} 
-   \frac{\theta_4^6(\tau)}{\eta^6(\tau)}  \Phi^{(r, s) }_{NS^-} 
\right]
\frac{E_{4,n}(q) }{\eta^8(\tau)}\otimes  \frac{ \Gamma_{2+n, 2}^{(r, s) } ( q, \bar q ) }{\eta^2(\tau)}.
\end{equation}
We define the lattice sum on $\Gamma_{2+n,2}^{r,s}$ as
\begin{eqnarray}\label{gamma2n}
\Gamma_{2+n, 2}^{(r, s)} (q, \bar q ) &= &
\sum_{\substack{m_1, m_2, n_2 \in \mathbb{Z}, \\ n_1 = \mathbb{Z} + \frac{r}{N} }}
q^{ \frac{p_L^2}{2} } \bar q^{\frac{p_R^2}{2}}  e^{2\pi i m_1 s/N} , \\ \nonumber
\frac{1}{2}{p_R^2}  &=& \frac{1}{2  T_2 U_2} | - m_1 U +m_2 + n_1 T + ( TU-\vec V^2) +b\cdot V|^2 , \\
\nonumber
\frac{1}{2}{p_L^2} &= &\frac{1}{2} p_R^2  + m_1 n_1 + m_2 n_2 +\frac{\vec b^2}{4} \,. 
\end{eqnarray}
where $T,U$ are Kahler and complex structure of the torus $T^2$.

The partition function on the $D6$ lattice in $R,NS^-, NS^+$ sectors are:
\begin{equation}
{\cal Z}_R ( D6; q ) = \frac{ \theta_2^6}{\eta^6}, \quad
{\cal Z}_{NS^+} ( D6; q ) = \frac{\theta_3^6}{\eta^6} , \quad
{\cal Z}_{NS^-} ( D6;  q) = \frac{ \theta_4^6}{\eta^6}. 
\end{equation}
$\theta_i$ where $i=2,3,4$ are standard Jacobi theta functions.
The trace on the $(6, 6)$ CFT is given by:
\begin{eqnarray} \label{d2k3part}
\Phi^{(r, s) }_R&=&\frac{1}{N} {\rm Tr}_{R\, R,g^r}[g^s(-1)^{F_R}q^{L_0-c/24}
\bar{q}^{\bar{L_0} -\bar{c}/24}],\\ \nonumber
\Phi^{(r, s) }_{NS^+} &=&\frac{1}{N} {\rm Tr}_{NS\, R,g^r}[g^s(-1)^{F_R}q^{L_0 -c/24}
\bar{q}^{\bar{L_0} -\bar{c}/24 }],\\ \nonumber
\Phi^{(r, s) }_{NS^-} &=&\frac{1}{N} {\rm Tr}_{NS\, R,g^r}[g^s(-1)^{F_R+F_L}
q^{L_0 -c/24 }\bar{q}^{\bar{L_0} -\bar{c}/{24} } ].
\end{eqnarray}
$\otimes$ refers to hatting procedure as in \cite{Weiss:2007tk,Datta:2015hza}. For a single Wilson line we have
\be
E_{4,1}=E_{4,1}^{even}\theta_{even}+E_{4,1}^{odd}\theta_{odd}
\ee
where $\theta_{even}(\tau)=\theta_{3}(2\tau,z)$ and $\theta_{odd}(\tau)=\theta_{2}(2\tau,z)$.
We have
\be
\hat E_{4,1}=E_{4,1}^{even}+E_{4,1}^{odd}.
\ee

In absence of any Wilson line the partition function from the second $E_8$ is given by $\frac{E_4(q) }{\eta^8(\tau)}$. If $n$ Wilson lines be present this will decompose the $E_4$ to its theta components accordingly and reduce the weight of the overall modular form by a factor of $n/2$. The resulting function $E_{4,n}$ will be of weight $4-n/2$.
\be
\hat E_{4,n}\in {\cal  M}_{4-n/2}
\ee
where ${\cal  M}_{4-n/2}$ is the space of modular forms of weight $4-n/2$.

The integrality of the Gopakumar-Vafa invariants at genus $g=0$ is satisfied if the following are integers:
\begin{eqnarray} \label{d2k3partint}
\sum_s e^{-2\pi i ks/N}\Phi^{(r, s)}_R,\quad \sum_s e^{-2\pi i ks/N}\Phi^{(r, s) }_{NS^+}, \quad \sum_s e^{-2\pi i ks/N}\Phi^{(r, s)}_{NS^-}
\end{eqnarray}
for every $k\in \mathbb{Z}$.
This extra phase $e^{-2\pi i ks/N}$ comes from $\Gamma_{2+n, 2}^{(r,s)}$ or its derivatives and will be clear from the next section \ref{gvextract}.
Using spectral flow symmetry and the definition of twisted elliptic genus \cite{Chattopadhyaya:2016xpa} we can say that
\begin{eqnarray} \label{relegenus}
\Phi^{(r,s)}_R &=&  F^{(r,s)}(\tau, \frac{1}{2} ), \\ \nonumber
\Phi^{(r,s)}_{NS^+} &=&  q^{1/4}  F^{(r,s)}(\tau,  \frac{\tau + 1}{2} ) , \\ \nonumber
\Phi^{(r,s)}_{NS^-} &=&  q^{1/4}  F^{(r,s)} ( \tau, \frac{\tau}{2} ) ,
\end{eqnarray}
where the twisted elliptic genus of $K3$ is defined as:
\begin{equation}\label{deftwistep}
F^{(r,s)}(\tau,z)=\frac{1}{N}{\rm Tr}_{R\, R g'^r}[(-1)^{F_{K3}+\bar{F}_{K3} }g'^s e^{2\pi i zF_{K3}}
q^{L_0-c/24} \bar{q}^{\bar{L_0}-\bar{c}/24}].
\end{equation}
Doing the $q$ expansion for any $z$ we have
\begin{eqnarray} \label{frsint}
\sum_{s=0}^{N-1}e^{-2\pi i ks/N}F^{(r,s)}(\tau,z)=\sum_{n,l} f(n,l) q^n z^l,
\end{eqnarray}
where $n\in \mathbb{Q}, l\mathbb{Z}$. A few examples are shown in Appendix \ref{lowcexp} and $f(n)$ can be seen to be integers.
In extracting the Gopakumar Vafa invariants we need the component of ${\cal Z}_{\rm new}$ given by:
\begin{equation} \label{decompk3part}
{\cal Z}_{\rm new}^{(r,s)} = 
\frac{1}{\eta^2(\tau) } 
\left [  \frac{\theta_2^6(\tau)}{\eta^6(\tau)}  \Phi^{(r, s) }_R  +
\frac{  \theta_3^6(\tau)}{\eta^6(\tau) } \Phi^{(r, s) }_{NS^+} 
-   \frac{\theta_4^6(\tau)}{\eta^6(\tau)}  \Phi^{(r, s) }_{NS^-} 
\right]
\frac{\hat E_{4,n}(q) }{\eta^8(\tau)}
\end{equation}
This shows that the only dependence on $s$ comes from which is $g'^s$ insertion on $K3$ while extracting the Gopakumar Vafa invariants.
Analyzing the $q$ expansions of the theta functions obtained from various lattice partition functions and the modular forms we can say that
the integrality of the Gopakumar Vafa invariants are ensured by the integrality of
the Fourier transform of the twisted elliptic genus as in (\ref{frsint}) for $z=\{\frac{1}{2},\frac{\tau}{2},\frac{\tau+1}{2}\}$ even though these modular forms do pick up some phases individually. We shall see in the next sub-section that these phases will eventually cancel and the integrality will solely depend upon the integer coefficients of the discrete Fourier transform of the twisted elliptic genus.

In terms of Eisenstein series the new-supersymmetric index can be given for standard embeddings as:
\begin{eqnarray} \label{zneweis}
{\cal Z}_{\rm new}(q, \bar q )  = \frac{1}{2\eta^{24} } \Gamma_{2,2}^{(r, s) } 
E_4  \left[ \frac{1}{4}  \alpha^{(r, s) }_{g'}  E_6  -   \beta^{(r, s) }_{g'}  E_4
\right].
\end{eqnarray}
The $\alpha^{(r, s) }_{g'}, \beta^{(r, s) }_{g'}$ are given in terms of the twisted elliptic genus as:
\begin{eqnarray} \label{twistellip}
F^{(0, 0)} (\tau, z) =   \alpha_{g'}^{(0,0)}   A(\tau, z) , \\ \nonumber
F^{(0,1)}( \tau, z) = \alpha_{g'}^{(0, 1) }   A (\tau, z)  + \beta_{g'}^{(0, 1) }(\tau)
B  ( \tau, z) ,
\end{eqnarray}
where  the Jacobi forms 
$A( \tau, z) $  and $B( \tau, z)$ are of weights 0 and $-2$ respectively and have index 1.
\subsection{Addition of Wilson lines}
Breaking of the $E_8$ gauge group with Wilson lines were studied in \cite{Ananthanarayan:1989ts,Weiss:2007tk} in details. 
In this section we shall review the hatting procedure for $n$ Wilson lines following the discussions in \cite{Weiss:2007tk}.

We can use the sequential Higgs mechanism and move along the moduli space away from the generic point. The $E_8$ can be sequentially broken to smaller subgroups. We shall review the method for one and two Wilson lines which breaks the $E_8$ to $E_7$ and $E_6$ respectively. The corresponding partition functions can be written down in terms of Jacobi theta functions. This corresponds to
putting constraints on the Wilson line moduli and reduces the number of free Wilson line moduli by 1 in each step. 

	\begin{figure}[H]
\begin{center}
	\includegraphics[width=10cm]{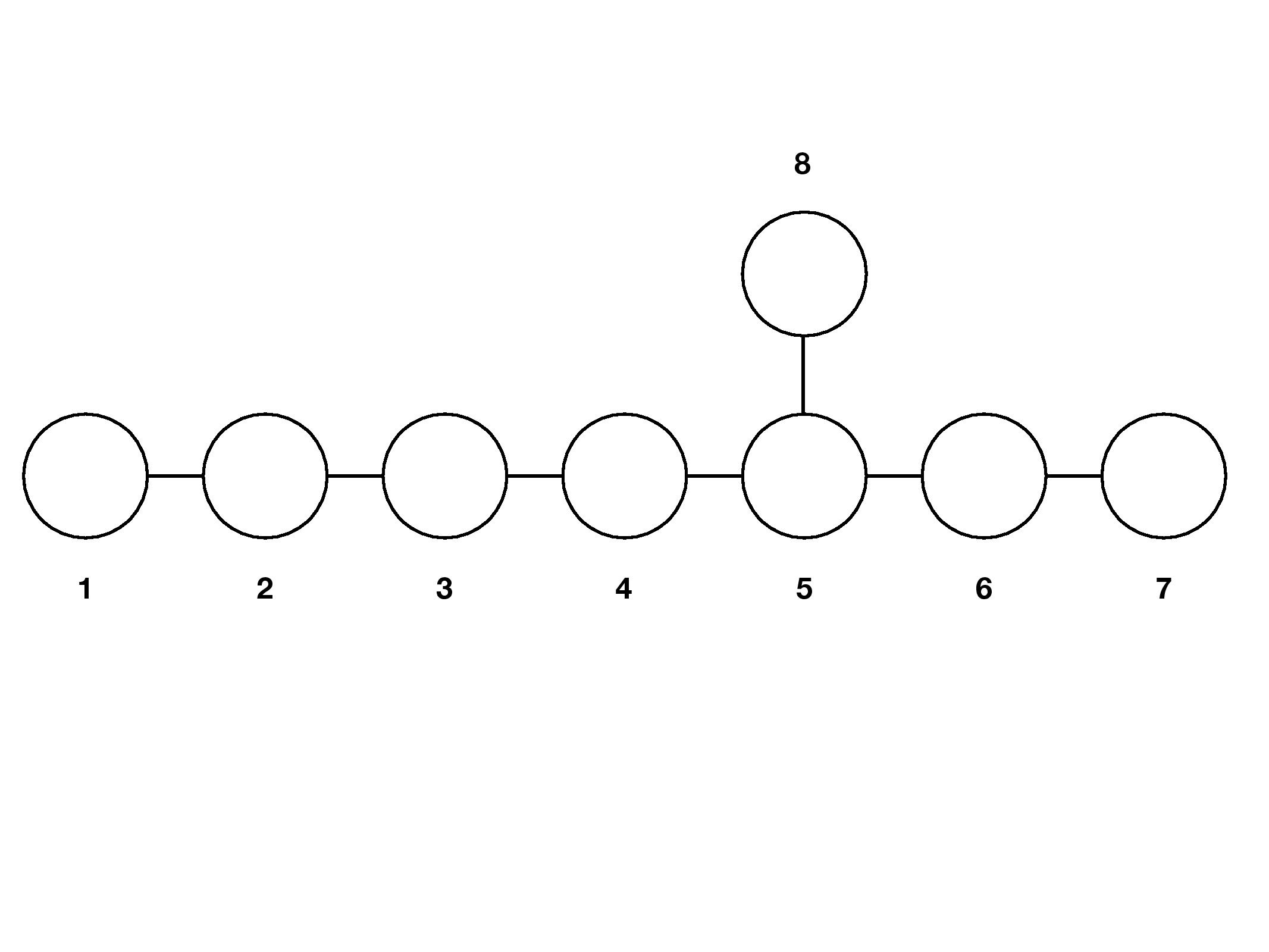}
	\caption{Dynkin diagram for $E_8$}
	\label{e8}
\end{center}
	\end{figure}

Assuming the roots  $\alpha_i$ are in positions $i$ in the above figure \ref{e8} we introduce a Wilson line moduli $z$, or for 7 Wilson lines we demand $\alpha_1\cdot z=0$.
In both cases we can write the $E_8$ partition function as:
\begin{eqnarray}
\sum_{p\in \Gamma_{E_8}} q^{p^2/2}&=&\sum_{n_i\in\mathbb{Z}}q^{n_1^2+n_2^2\cdots n_8^2-n_1n_2-n_2n_3-n_3n_4-n_4n_5-n_5n_6-n_6n_7-n_5n_8}\\ \nn
&=& \sum_{n_i\in\mathbb{Z}}q^{(n_1-n_2/2)^2+\frac{3n_2^2}{4}\cdots n_8^2-n_1n_2-n_2n_3-n_3n_4-n_4n_5-n_5n_6-n_6n_7-n_5n_8} \\ \nn
&=&\sum_{j=0}^1 \sum_{n_1\in\mathbb{Z}}q^{(n_1-j/2)^2}\sum_{n_2,\cdots n_8\in \mathbb{Z}} q^{\frac{3}{4}(2n_2-j)^2+n_3^2\cdots n_8^2-(2n_2-j)n_3-n_3n_4-\cdots n_6n_7-n_5n_8}\\ \nn
&=& \theta \left[\begin{smallmatrix}
j/2\\ 0
\end{smallmatrix}\right](2\tau) \sum_{n_2,\cdots n_8\in \mathbb{Z}} q^{\frac{3}{4}(2n_2-j)^2+n_3^2\cdots n_8^2-(2n_2-j)n_3-n_3n_4-\cdots n_6n_7-n_5n_8}\\ \nn
\end{eqnarray}
The second sum can also be given in terms of Jacobi theta functions as:
\begin{eqnarray}
\sum_{p\in E_7^{(1)}}q^{p^2/2} &=& \sum_{n_2,\cdots n_8\in \mathbb{Z}} q^{\frac{3}{4}(2n_2-j)^2+n_3^2\cdots n_8^2-(2n_2-j)n_3-n_3n_4-\cdots n_6n_7-n_5n_8}\\ \nn
&=& \sum_{\{a,b\}\in \{0,1\}}\theta  \left[\begin{smallmatrix}
a/2+j/2\\ 0
\end{smallmatrix}\right](2\tau)\;\;\theta \left[\begin{smallmatrix}
a/2\\ b/2
\end{smallmatrix}\right]^6(-1)^{jb}.
\end{eqnarray}
The above splitting works for both 1 and 7 Wilson line moduli being added and the lattice $E_8\supset E_7\times SU(2)$.

Now for the breaking of $E_8 \supset E_6\times SU(3)$ we have the following:
\begin{eqnarray}\nn
\sum_{p\in \Gamma_{E_8}} q^{p^2/2}&=&\sum_{n_1,n_2\in\mathbb{Z}}q^{(n_1-j_1/2)^2+3(n_2+j_1/2-j_2/3)^2}\sum_{n_3\cdots n_8\in\mathbb{Z}} q^{\frac{2}{3}(3n_3-j_2)^2+n_4^2\cdots n_8^2-3(n_3-j_2)n_4-n_4n_5-\cdots} \\ \nn
&=&\sum_{j_1=0,1}\sum_{j_2=0,1,2}\theta \left[\begin{smallmatrix}
j_1/2\\ 0
\end{smallmatrix}\right](2\tau)\;\;\theta \left[\begin{smallmatrix}
j_1/2+j_2/3\\ 0
\end{smallmatrix}\right](6\tau) \sum_{\{a,b\}\in \{0,1\}} \theta \left[\begin{smallmatrix}
a/2+j_2/3\\ 0
\end{smallmatrix}\right](3\tau)\;\;\theta \left[\begin{smallmatrix}
a/2\\ b/2
\end{smallmatrix}\right]^5(-1)^{j_2b}.\\
\end{eqnarray}

In general we can break the $E_8$ lattice similarly to other smaller rank gauge groups by adding more Wilson lines. We can have lattice decompositions with 3 or 5 Wilson lines as $E_8\supset SO(10)\times SU(4)$ and with 4 Wilson lines as $E_8\supset SU(5)\times SU(5)$.

%For a $\le 4$ Wilson lines $STUV$ model the hatting procedure would be given by writing the modular form associated with the $E_8$ lattice.
%\begin{eqnarray}
%\sum_{p\in \Gamma_{E_8}} q^{p^2/2}&=& f_0^k \theta_0^{(8-k)}+\cdots f_k^k \theta_k^{(8-k)}
%\end{eqnarray}
%where,

In the single Wilson line case or the
STUV model, the relevant modular forms are Jacobi forms of index 1.
We write this as $f(\tau,z)$ which admits a theta decomposition:
\be
f(\tau,z)=\hat f^{odd}(\tau)\theta_{odd}(\tau,z)+\hat f^{even}(\tau)\theta_{even}(\tau,z).
\ee
where
\be\nn
\theta_{odd}(\tau,z)=\theta_2(2\tau,2z),\qquad \theta_{2even}(\tau,z)=\theta_3(2\tau,2z).
\ee
For a generalized hatting procedure,
for $n\le 4$ Wilson lines we have
\be
\hat f\left[\begin{smallmatrix}
	a\\b
\end{smallmatrix}\right](\tau,V_1,\cdots V_n)=\hat f^n_0\left[\begin{smallmatrix}
a\\b
\end{smallmatrix}\right](\tau)+\cdots \hat f^n_n\left[\begin{smallmatrix}
a\\b
\end{smallmatrix}\right](\tau)
\ee

From the analysis of \cite{Weiss:2007tk} for when $n$ is even or odd, the lattice split differs as follows. For odd $n$ we have the following hatting
\begin{eqnarray}
\hat f_J^n=\sum_{J=0}^{n}\sum_{a,b=\{0,1\}}\theta\left[\begin{smallmatrix}
a/2+J/(n+1)\\
0
\end{smallmatrix}\right]((n+1)\tau) \theta\left[\begin{smallmatrix}
a/2\\
b/2
\end{smallmatrix}\right]^{7-n} (-1)^{bJ}
\end{eqnarray}
where,
\be
\theta\left[\begin{smallmatrix}
	a\\
	b
\end{smallmatrix}\right](\tau)=\sum_{n\in\mathbb{Z}}q^{(n-a)^2/2}e^{-2\pi i b(n-a)}
\ee
%Sometimes the explicit dependence of $\tau$ is omitted when there is numerical factor with it.
We can see that $q$ expansion of the above function $\hat f_J^k$ will have integer coefficients.

For the even number of Wilson lines ie, $n\in 2\mathbb{Z}$ we can write
\begin{eqnarray}
\hat f_J^n=\sum_{J=0}^{k}\sum_{a,b=\{0,1\}}\theta\left[\begin{smallmatrix}
a/2+J/(n+1)\\
b/2
\end{smallmatrix}\right]((n+1)\tau)\;\; \theta\left[\begin{smallmatrix}
a/2\\
b/2
\end{smallmatrix}\right]^{7-n} (-1)^{bJ}
\end{eqnarray}
Now if we carry out the sum over $J$ different phases will cancel and integers will be picked up.
%\footnote{I am not fully confident on this yet, whether this sum is allowed without taking the $\Gamma_{2+n, 2}^{r,s}$ factor, most likely it is, because then we just pick up the $\delta$ function condition} 

When the number of Wilson lines is $n>4$ we need the complementary part of the relevant theta decomposition, this is given by,
\be
\check f\left[\begin{smallmatrix}
	a\\b
\end{smallmatrix}\right](\tau,V_1,\cdots V_n)=\theta^{(8-n)}_0\left[\begin{smallmatrix}
	a\\b
\end{smallmatrix}\right](\tau)+\cdots  \theta^{(8-n)}_n \left[\begin{smallmatrix}
	a\\b
\end{smallmatrix}\right](\tau)
\ee
where we have,
\begin{eqnarray}\nn
\theta^{(k)}_J&=&\sum_{\begin{smallmatrix}
	j_1=0,1\\ \cdots \\ j_{k-1}=0,1,\cdots k-1
	\end{smallmatrix}}\theta\left[\begin{smallmatrix}
j_1/2\\0
\end{smallmatrix}\right](2\tau)\;\; \theta\left[\begin{smallmatrix}
j_2/3-j_1/2\\0
\end{smallmatrix}\right](6\tau)\cdots \theta\left[\begin{smallmatrix}
\frac{j_{k-1}}{k}-\frac{j_{k-2}}{(k-1)}\\0
\end{smallmatrix}\right](k(k-1)\tau)\\
&& \theta\left[\begin{smallmatrix}
\frac{J}{(k+1)}-\frac{j_{k-1}}{k}\\0
\end{smallmatrix}\right](k(k+1)\tau)
\end{eqnarray}

For a general $n$ the associated modular form from the $E_8$ lattice which appears in the new supersymmetric index would be given by,
\begin{eqnarray}
E_{4,n}&:=&\hat f\left[\begin{smallmatrix}
	a\\b
\end{smallmatrix}\right](\tau,V_1,\cdots V_n)\qquad {\rm for \; n\le 4};\\ \nn
&&\check f\left[\begin{smallmatrix}
a\\b
\end{smallmatrix}\right](\tau,V_1,\cdots V_n) \qquad {\rm for \; n>4}
\end{eqnarray}

Interestingly $E_{4,8}$ is just 1 and the second lattice is completely removed by the Wilson line addition and for $K3\times T^2$ this was discussed in \cite{Harvey-Moore}.

\section{Gopakumar-Vafa invariants from gravitational threshold}\label{gvextract}
In this section we briefly describe the method of extracting the Gopakumar Vafa invariants of the dual Calabi Yau threefolds from a one loop gravitational coupling computation n the heterotic side using the new super-symmetric index.

The gravitational coupling $F_g$ appear in the low energy theories of the heterotic ${\cal N}=2$ compactifications as:
\begin{equation}
S = \int F_g(y,\bar y) F_+^{2g-2} R_+^2 ,
\end{equation}
Here $F_+,R_+$ denote the self dual parts of the graviphoton and the Riemann tensor. For $K3\times T^2$ this was initially computed in \cite{Marino:1998pg} and for additional Wilson lines this was computed in \cite{Weiss:2007tk}.

From the  world sheet analysis of \cite{Antoniadis:1995zn} we see that the one loop integral is given by 
\begin{eqnarray} \label{fgexp}
F_g &=& \frac{1}{ 2\pi^2( g!) ^2} \int \frac{ d^2 \tau}{\tau_2} \left\{  \frac{1}{\tau_2^2 \eta^2(\tau)}
{\rm Tr}_R
\left[   ( i \bar \partial X)^ {( 2g - 2) } 
(-1)^{F} F q^{L_0 - \frac{c}{24} } \bar q^{\tilde{L}_0 - \frac{\tilde{c}}{24} } \right ]  \right. \\ \nonumber
&& \qquad \qquad \qquad  \left.  \langle \prod_{i=1}^g \int d^2 x_i Z^1 \partial Z^2 (x_i)  
\prod_{j = 1}^g \int d^2 y_j \bar Z^2  \partial \bar Z^1 ( y_j)  \rangle  \right\}.
\end{eqnarray}
where $X$ is the  complex coordinate on the torus $T^2$; $Z^1, Z^2$ are the complex
coordinates of the transverse non-compact bosons \cite{Antoniadis:1995zn}. The trace part corresponds to the new super-symmetric index with some derivatives acting on the Siegel Narain theta functions coming from the  bosonic zero modes on $T^2$.

 In this work we shall generalize this to several orbifolds of $K3\times T^2$ where the orbifold action is an action $g'$ on $K3$ which corresponds to $[M_{23}]$ group together with a $1/N$ shift on one circle of $T^2$, in presence of Wilson lines.

A similar computation was done in our earlier work \cite{Chattopadhyaya:2017zul} and generalized to some non-standard embeddings in \cite{Banlaki:2018pcc,Banlaki:2019bxr} but in absence of any Wilson line in the theory ($V=0$).
Using the gravitational couplings $F_g$ we can extract the data of the Calabi Yau geometry corresponding to the dual type IIA picture.
The most important ingredient in computing the $F_g$ is given by the new super-symmetric index and its Fourier coefficients in the $q$ expansion ensures the possibility of integer Gopakumar Vafa invariants.
%This we shall discuss for generic cases in this section.

In absence of any Wilson line the purely holomorphic contribution to the gravitational couplings which can be directly related to the Gopakumar Vafa invariants in the dual Calabi Yau geometry are given by
\bea\label{fhol}\nonumber
\bar{F}_g^{{\rm hol}}&=&\frac{(-1)^{g-1}}{\pi^2} \sum_{s=0}^{N-1} \left( \sum_{m>0}e^{-2\pi i n_2s/N}c^{(r,s)}_{g-1}(m^2/2){\rm Li}_{3-2g}(e^{2\pi im\cdot y})+\frac{1}{2}c_{g-1}^{(0,s)}(0)\zeta(3-2g)
\right), \\
\eea
and the coefficients are given by
\begin{equation}\label{defcg}
{\cal Z}_{\rm new}^{r,s}\tilde{\cal P}_{2g} (\tau) 
=\sum_{l \in \frac{\mathbb{Z}}{N} } c_{g-1}^{(r,s)} (l) q^l .
\end{equation}
where $\tilde{\cal P}_{2g}$ is ${\cal P}_{2g}$ evaluated at $\tau_2\rightarrow\infty$
and
\begin{equation}\label{relfgfhol}
F_g^{{\rm GV}}=\frac{(-1)^{g+1}}{2(2\pi )^{2g-2}} \bar F_g^{\rm hol}.
\end{equation}
where the LHS denotes the generating function for genus $g$ Gopakumar Vafa invariants.
\bea
F^{{\rm GV}}_g&=&\frac{(-1)^g |B_{2g}B_{2g-2}|\chi(X)}{4g (2g-2)(2g-2)!}\\ \nn
&+& \sum_{\beta}\left[\frac{|B_{2g}|n_m^0}{2g (2g-2)!}+\frac{2(-1)^g n_m^2}{(2g-2)!}\pm...-\frac{g-2}{12}n_{m}^{g-1}+n_m^g\right]{\rm Li}_{3-2g}(e^{2\pi i m\cdot y}).
\eea
The sum over $m>0$ will be discussed in the next subsection in more details.

The calculation of $F_g$ and extraction of its holomorphic component can be done following the unfolding technique as in \cite{Dixon:1990pc,Harvey-Moore,Borcherds:1996uda,Chattopadhyaya:2017zul}. In the following two sections we shall mainly require to adapt these methods for different components that constitute $F_g$. The integral is carried out by unfolding the fundamental domain to the whole upper half plane and involves three orbits 
\begin{enumerate}
	\item Zero orbit (only contributes to $g=1$),
	\item Non-degenerate orbit
	\item Degenerate orbit.
\end{enumerate}
The evaluation of the integral in non-degenerate orbit is crucial in obtaining the Gopakumar Vafa invariants.

\subsection{Non-degenerate orbit of the integral for $F_g$}\label{nondeg}
In this section we shall give more details of the integral corresponding to non-degenerate orbit.

\begin{eqnarray}
F_g(T, U,\vec  V) = \int_{{\cal F}} \frac{d^2\tau}{\tau_2} \tau_2^{2g-2} 
\sum_{\gamma,r,s} {\cal Z}_{{\rm new}}(r,s) ( p_R^{(r,s)} )^{2g-2} q^{|p_L|^2/2}\bar q^{|p_R|^2/2}{\cal P}_{2g}.
\end{eqnarray}
${\cal F}$ is the fundamental domain of $SL(2,\mathbb{Z})$, $P_{2g}$ is a weakly holomorphic modular form and for lower values of $g$ it is given by,
\begin{eqnarray}
{\cal P}_0 &=&-1,\\ \nn
{\cal P}_2&=& - \frac{\pi^2}{3} \hat E_2,\\ \nn
{\cal P}_4  &= &- \pi^4 \left( \frac{1}{18} \hat E_2^2 + \frac{1}{90} E_4 \right),
\end{eqnarray}
where $p_L,p_R$ are given in equation (\ref{gamma2n}), $\hat E_2=E_2-\frac{3}{\pi\tau_2}$.

 We need to do the following integration:
\begin{eqnarray}
{\cal I} = \int_{{\cal F}} \frac{d^2\tau}{\tau_2} \tau_2^{2g-t-2} 
\sum_{\gamma} ( p_R ^{(r,s)})^{2g-2} q^{\frac{|p_L|^2}{2}}\bar q^{\frac{|p_R|^2}{2}}  {\cal Z}_{\rm new}^{(r,s)},
\end{eqnarray}
We define $Y=\det\;{\rm Im}\left(\begin{matrix}U & \vec V\\ \vec V & T \end{matrix}\right)$.
This gives for a generic term which extracts $q^{n'}$ from the new super-symmetric index,
\begin{eqnarray}\nn
{\cal I^{\rm generic}}&=&\int_{{\cal F}} \frac{d^2\tau}{\tau_2}\frac{1}{(2Y)^{g-1}}(-m_1U+n_1T+ n_2(TU-\vec V^2)+  m_2+ \vec b\cdot \vec V)^{2g-2}\\
&&\tau_2^{2g-t-2} \exp\left(2\pi i\tau(m_1n_1+(m_2n_2+\frac{1}{4}b^2))\right) q^{n'}\\ \nn
&& \exp\left(-\frac{\pi\tau_2}{Y}|(-m_1U+n_1T+ n_2(TU-\vec V^2)+ m_2+ \vec b\cdot \vec V)|^2 \right)
\end{eqnarray}

We now perform the Poisson re-summation over the variables $(m_1, m_2)$ in $p_L,p_R$ to re-write the integrand as:
\be \label{igen}
{\cal I^{\rm generic}}=\int_{\cal F}\frac{Y}{U_2} \frac{d^2\tau}{\tau_2^2} \left(\frac{\sqrt{Y}}{\sqrt{2}U_2}\right)^{2g-2}(n_1\tau+k_1+U k_2+(TU-\vec V^2) n_2\tau)^{2g-2}\,e^{\cal G}.
\ee
In the above equation we have,
\begin{eqnarray}\label{eg}
{\cal G}&=&-\frac{\pi Y}{U_2^2 \tau_2}|{\cal A}|^2-2\pi i \det A+\frac{\pi \vec b}{U_2}\cdot (\vec V \tilde {\cal A}-\vec {\bar V}{\cal A})\\ \nn
&-&\frac{\pi n_2}{U_2}(\vec V_2^2\tilde{\cal A}-\vec{\bar V}_2^2 {\cal A}) +2\pi i \frac{\vec V_2^2}{U_2^2}(n_1+n_2 \bar U){\cal A}+2\pi i \tau \frac{\vec b^2}{4}+2\pi i\tau n'\\ \nn
A &=&\left(\begin{matrix}
n_1 & k_1 \\ n_2 & k_2
\end{matrix}\right)\\ \nn
{\cal A} &=& \left(\begin{matrix}
1 & U
\end{matrix}\right) A\left(\begin{matrix}
\tau \\ 1 
\end{matrix}\right)\\ \nn
\tilde{\cal A} &=& \left(\begin{matrix}
1 & \bar U
\end{matrix}\right) A\left(\begin{matrix}
\tau \\ 1 
\end{matrix}\right).
\end{eqnarray}
In the non-degenerate orbit, $n_2=0,k_2\in \mathbb{Z}$ and $k_1\in \mathbb{Z}+s/N$ and $0\le k_1<n_1$ and $k_1>0$. In the twisted sector $r$ we have $0\le k_1<N n_1+r$, $n_1\in \mathbb{Z}+r/N$ while $n_2=0,k_2\in \mathbb{Z}$ remain as in the untwisted sector.
The details of doing similar integrals have been discussed before in the literature quite extensively \cite{Dixon:1990pc,Harvey-Moore,Borcherds:1996uda,David:2006ji,Chattopadhyaya:2017zul}.
In the non-degenerate orbit we take
\[
A =\left(\begin{matrix}
	n_1 & k_1 \\ 0 & p
\end{matrix}\right).\]
In the untwisted sector when $n_1>0$ we have the result as:
\begin{eqnarray}\nn
{\cal I}_{n_1>0}&=& 2^{2-g} e^{-2\pi i n_2s/N}c^{(r,s)}(n_1n_2-\vec b^2/4)\sum_{h=0}^{2g-2}\sum_{j=0}^{[g-1-h/2]}\frac{(2k)!}{j! h^!(2g-2-h-2j)!( 4\pi) ^j}\\ \nn
&&p^{h} (Y )^{j+h-1/2-g} (n_2 U_2+n_1 T_2+\vec b \cdot \vec V_2)^{2g-2-h-2j} (-1)^{k-j}\\ \nn
&& \left( \frac{|p|Y}{n_1 T_2+n_2 U_2+\vec b \cdot \vec V_2} \right)^{2g-2-h-j-t-1/2}\\ 
&& K_{2g-2-h-j-t-1/2}(2\pi |p|(n_1T_2+n_2 U_2+\vec b \cdot \vec V_2)) e^{-2\pi i p(n_1T_1+n_2 U_1+\vec b \cdot \vec V_1)},
\end{eqnarray}
where $n'$ in equation (\ref{eg}) is related as $n'=4n_1n_2-\vec b^2$.
The phase $e^{-2\pi i n_2s/N}$ comes from the integral on $\tau_1$ which becomes a Kronecker delta function for $k_1$ dependence \cite{David:2006ji,Chattopadhyaya:2017zul}. The sum on $n_1,n_2,r,s,b$ are implied in the above equation.
In general we can write the $\vec b^2$ as follows:

\begin{table}[H]
	\renewcommand{\arraystretch}{0.5}
	\begin{center}
		\vspace{0.5cm}
		\setlength{\doublerulesep}{20\arrayrulewidth}
		{\scriptsize{
				\begin{tabular}{|c|c|}
					\hline
					&  \\
					$n$ & $b^2/4$\\
					\hline 
					& \\
					0 & 0\\ 
					1 & $\frac{b_1^2}{4}$ \\
					2 & $b_1^2-b_1 b_2+\frac{b_2^2}{3}$\\
					3 & $b_1^2+b_2^2-b_1 b_2-b_2b_3+\frac{3b_3^2}{8}$\\
					4 & $b_1^2+b_2^2+b_3^2-b_1 b_2-b_2b_3-b_3b_4+\frac{2b_4^2}{5}$\\
					5 & $\frac{5b_4^2}{8}+b_5^2+b_6^2+b_7^2+b_8^2-b_4b_5-b_5b_6-b_5b_8-b_6b_7-b_7b_8$ \\
					6 & $\frac{2b_3^2}{3}+b_4^2+b_5^2+b_6^2+b_7^2+b_8^2-b_3b_4-b_4b_5-b_5b_6-b_5b_8-b_6b_7-b_7b_8$ \\
					7 & $\frac{3b_2^2}{4}+b_3^2+b_4^2+b_5^2+b_6^2+b_7^2+b_8^2-b_2b_3-b_3b_4-b_4b_5-b_5b_6-b_5b_8-b_6b_7-b_7b_8$ \\
					
					\hline
				\end{tabular}
		}}
		
	\end{center}
	\vspace{-0.2cm}
	\caption{Norm $b^2$ for $n$ Wilson lines}
	\renewcommand{\arraystretch}{0.5}
\end{table}

We can re-write the above Bessel function with poly-logarithms to get to a form similar to the Calabi Yau duals.
To do so we use the relations:
\begin{eqnarray} \label{polylog}
K_{s+1/2}(x)&=&\sqrt{\frac{\pi}{2x}}e^{-x}\sum_{k=0}^{s}\frac{(s+k)!}{k!(s-k)!}\frac{1}{(2x)^k},\\
{\rm Li}_m(x)&=&\sum_{l=1}^\infty \frac{x^l}{l^m}, 
\end{eqnarray}
where $l=1,2,...\infty$. 
%In this case $p=l/\zeta_J^{1/2}$ with $l\in \mathbb{Z}$.
This leads to 
\begin{eqnarray} \nn
{\cal I}_{n_1>0}&=& 2^{2-g} \sum_{h=0}^{2g-2}\sum_{j=0}^{[g-1-h/2]}\sum_{a=0}^s\frac{(2g-2)!}{j! h^+!(2g-2-h-2j)!}\frac{(s+a)!}{a!(s-a)!}\frac{(-1)^{g-1-j+h}}{( 4\pi) ^{j+a}}\\ \nn
&& (Y )^{g-1-t} {\rm(sgn\;Im}(m\cdot y ))^{h}
{\rm Im}(m\hat{\cdot}y)^{t-j-a}{\rm Li}_{3+a+j+t-2g}(e^{2\pi i m\hat{\cdot}y}).\\ \label{resultli}
\end{eqnarray}
where, $m\cdot y = n_1T+n_2U+\vec b\cdot \vec V$ and \[m\hat{\cdot}y={\rm Re}(n_1T+n_2U+\vec b\cdot V)+|{\rm Im}(n_1T+n_2U+\vec b\cdot V)|.\]
The contribution to the coupling non-degenerate orbit would be given by,
\begin{eqnarray} \label{fresultfg}
{\cal I}^{\rm nondeg}_{g>1} &=& \frac{(-1)^{g-1}}{2^{2(g-1)}\pi^2} \sum_{h=0}^{2g-2}\sum_{j=0}^{g-1-h/2}\sum_{a=0}^{\hat s} \frac{(2g-2)!}{j!h!(2g-h-2j-2)!}\\ \nn &&\frac{(-1)^{j+h}}{(4\pi)^{j+a}}\frac{(\hat s+a)!}{(\hat s-a)!a!}({\rm sgn}({\rm{Im} }(m\cdot y)))^h  \\ \nn
&&\frac{({\rm{Im}}(m\hat \cdot y))^{t-j-a}}{(Y)^t} {\rm Li}_{3+a+j+t-2g}(e^{2\pi i m\hat \cdot y}) e^{-2\pi i n_2s/N}c_{g-1}^{(r,s)}(n_1n_2-\vec b^2/4,t)\\ \nn
%&+& \sum_{s=0}^{N-1} \,\frac{c^{(0,s)}_{g-1}(0,t)}{2^{4g-5}\, \pi^{g+1}}\frac{1}{(Y)^{g-1}} \sum_{\tilde s=0}^{g-1}(-1)^{\tilde s} \frac{(2g-2)!}{\tilde{s}!(g-1-\tilde s)!}\psi (1/2+\tilde {s})\\ \nn
&+& \sum_{s=0}^{N-1}\sum_{t=0}^{g-2}\,\frac{c^{(0,s)}(0,t)}{\pi^{t+5/2}} \, \frac{\zeta(3+2(t-g))}{(Y)^t}
\times \\ 
&& \sum_{\tilde s=0}^{g-1}(-1)^{\tilde s} 2^{2(\tilde s-2g+2)}\frac{(2g-2)!}{(2\tilde s)!(g-1-\tilde s)!}\Gamma(3/2+\tilde s+t-g). 
\end{eqnarray}
where,
\begin{equation}\label{defcgtau}
{\cal Z}_{\rm new}^{(r, s)} ( \tau) {\cal P}_{2k +2} (\tau)  = 
\sum_{l \in \frac{\mathbb{Z}}{N} , t = 0 }^{t = g} c_{g-1}^{(r,s)} ( l, t) \tau_2^{-t} q^l .
\end{equation}
Here 
\begin{equation}\label{hatsdef}
\hat s + 1/2 = |\nu|, \qquad \qquad |\nu| = 2g - h - j - t -5/2.
\end{equation}
The last two lines in the equation (\ref{fresultfg}) comes from evaluating integral where $n_1=0$. This result can be extrapolated to $g=0$ to evaluate the Gopakumar Vafa invariants and the Euler character $\chi$ of the dual Calabi Yau geometry.

To evaluate the terms which do not mix the holomorphic and anti-holomorphic components we get $a=t=j=h=0$ terms and these are obtained from the anti-holomorphic piece $n_1,n_2<0$.
\bea
{\bar F_g}^{\rm hol}&=&\sum_{m>0}\frac{(-1)^{g-1}}{2^{2(g-1)}\pi^2}\sum_{s=0}^{N-1}\left({\rm Li}_{3-2g}(e^{2\pi i m\hat \cdot y}) e^{-2\pi i n_2s/N}c_{g-1}^{(r,s)}(n_1n_2-\vec b^2/4)\right.\\ \nn
&&+\left.\frac{1}{2}c_{g-1}^{(0,s)}(0,0)\zeta(3-2g)\right)
\eea
Here $m>0$ refers   to
$(n_1, n_2,\vec b)$, in the chamber $|\vec b\cdot \vec V_2|<<U_2<T_2/N,\\
n_1 \in {\mathbb{Z}}+{r/N}, n_2\in \mathbb{Z}$ with the restrictions
\begin{eqnarray}\label{restrict}
n_1, n_2 \geq 0, \quad\hbox{but } (n_1, n_2) \neq ( 0, 0 ), \\ \nonumber
( r/N, -n_2) , \qquad \hbox{with}\;\;  n_2>0 \;\;\hbox{and}\;\; r n_2 \leq N.
\end{eqnarray}

\noindent
We can now extract the Gopakumar-Vafa invariants from ${\bar F_g}^{\rm hol}$ by removing the factor 
\[\frac{(-1)^{g+1}}{2(2\pi)^{2g-2}}\].
Lists of Gopakumar Vafa invariants corresponding to standard and non-standard embeddings for order 2 and 3 orbifolds are listed in Appendix \ref{gv}.

\section{Perturbative pre-potential}\label{prepot}
The perturbative heterotic pre-potential in STUV model for the $K3\times T^2$ is given as
\be
F_0^{\rm het}=-S(TU-V^2)+p_n(T,U,V)-\frac{1}{4\pi^3}\sum_{n_1,n_2,b>0}c(n_1n_2-b^2/4){\rm Li}_3(e^{2\pi i(n_1T+n_2U+bV)}).
\ee
$p_n(T,U,V)$ is a polynomial of degree 3 in $T,U,V$.
It is possible to extract the coefficients of the degree 3 polynomial from $F_1$ computing the zero and degenerate orbit.
In this section we compute the perturbative polynomial from the heterotic string compactified on $K3\times T^2$ and two $E_8$ lattices with a $g'$ automorphism on $K3$ and $1/N$ shift alone one circle of $T^2$. This calculation is encoded in one loop correction of the gravitational coupling with one Wilson line applied to the $E_8$ lattices.
We follow the technique of Harvey-Moore in \cite{Harvey-Moore} to determine the triple intersection numbers using the symmetric polynomial of the heterotic pre-potential. In this process we need to integrate over the degenerate orbit and zero orbit in order to get the polynomial term for $T,U,V$ at genus 1. In this section we shall sketch the process and where it varies from the original case of $K3\times T^2$ due to the presence of the orbifold actions.

The computation goes very similarly  as described in details in \cite{Chattopadhyaya:2017zul}. First we need the new-super-symmetric index ${\cal Z}_{{\rm new}}$ for computing the $F_g$ in the degenerate orbit for $g=1$. This was computed for several standard embeddings (equation (\ref{zneweis})) and also some non-standard embeddings in \cite{Chattopadhyaya:2016xpa} and also recently in \cite{Banlaki:2019bxr}.

To evaluate the degree three polynomial 
we need to evaluate the following integration:
\begin{eqnarray}
F_1(T, U, V) = \int_{{\cal F}} \frac{d^2\tau}{\tau_2} 
 {\cal Z}_{{\rm new}}  {\cal P}_2.
\end{eqnarray}
The result from the zero and degenerate orbit will be important in getting the polynomial term and triple intersection numbers of the dual Calabi Yau.

Let us describe the computation for $K3$ which will also be required for the $(0,0)$ sector of the new supersymmetric index. We write ${\cal Z}_{\rm new}$ to be $\frac{E_4E_6}{\eta^{24}}\Gamma_{2,2}$ and $ {\cal P}_2=E_2-\frac{3}{\pi\tau_2}$ in the absence of Wilson lines.
If we add one Wilson line, for different embeddings we get the following result for 
${\cal Z}_{\rm new}$ to be 
\be
\frac{(12-n)\hat E_{4,1}E_6+(12+n)\hat E_{6,1}E_4}{24\eta^{24}} \otimes \Gamma_{3,2},
\ee
where $\otimes$ carries out the theta decomposition of the relevant Jacobi forms (in this case even and odd).

In the zero orbit \cite{Harvey-Moore,Lerche:1988np} the contribution only comes from $A=0$ which appears in the $(r,s)=(0,0)$ sector and is be given by
\be
{\cal I}_0=\frac{\pi Y}{6 U_2}E_2^2 F(q)|_{q^0}
\ee
$F(q)$ for each individual cases can be given only from $b^2=0,1,4$ terms.
Since the $(0,0)$ sector is given by $1/N$ of that of $K3$, so
${\cal I}_0=\frac{1}{N}{\cal I}_0^{K3}$.

In the STUV model we have
\be
F(q)(E_2-\frac{3}{\pi\tau_2})=\sum_{n\in\mathbb{Z}-b^2/4}\tilde{c}(n-b^2/4)q^n-\frac{3}{\pi\tau_2}\sum_{n\in\mathbb{Z}-b^2/4} c(n-b^2/4)q^{n-b^2/4}
\ee
where $b=0,\pm 1,\pm 2$ or $b>0, b\in \mathbb{Z}$.

The degenerate orbit is described by $A=\left(\begin{matrix}
0 & j\\ 0 & p
\end{matrix}\right)$. Here for genus one it is easier to consider $j,p\ne 0$.
After evaluating the $\tau_1$ integral the sum on $j$ can be performed
using a Sommerfeld-Watson transformation given by,
\begin{eqnarray}
\sum_{j=-\infty}^{\infty} \frac{e^{i\theta j}}{(j+B)^2+C^2} &=& \frac{\pi}{C}\frac{e^{-i\theta(B-iC)}}{1-e^{-2\pi i (B-iC)}}\\ \nn
&+& \frac{\pi}{C}{e^{-i\theta(B+iC)}}\frac{e^{2\pi i (B+iC)}}{1-e^{-2\pi i (B+iC)}}\\ \nn
\sum_{j=-\infty}^{\infty} \frac{e^{i\theta j}}{((j+B)^2+C^2)^2} &=& -\frac{1}{2C}\partial_C \sum_{j=-\infty}^{\infty} \frac{e^{i\theta j}}{(j+B)^2+C^2}\\ \nn
&& {\rm C>0,\qquad 0<\theta<2\pi}.
\end{eqnarray}
The sums on $p$ would result in poly-logarithm terms with odd order and these would give the genus 1 Gopakumar Vafa invariants at $n_1=0$.
The $p=0$ term however would result in poly-logarithms of even order and at $\theta=0$ these are given by even zeta functions. For $g=1$ we shall only require ${\rm Li}_2$ and ${\rm Li}_4$. The compact expressions for these are given by,
\begin{eqnarray}
{\rm Re}({\rm Li}_2(e^{i\theta}))=\sum_{j=1}^{\infty}\frac{\cos(\theta j)}{j^2}=\frac{\pi^2}{6}+\frac{\theta(-2\pi+\theta)}{4} \\ \nn
{\rm Re}({\rm Li}_4(e^{i\theta}))=\sum_{j=1}^{\infty}\frac{\cos(\theta j)}{j^4}=\frac{\pi^4}{90}-\frac{\theta^2(2\pi-\theta)^2}{48}
\end{eqnarray}
The explicit results 
from the degenerate orbit would be given by,
\begin{eqnarray}\label{degk3}
{\cal I}_{\rm deg}&=&\tilde{c}(0)\frac{\pi}{3}U_2-c(0)\frac{\pi}{15} \frac{U_2^3}{Y}-\frac{3}{\pi^2 Y}c(0)\zeta(3)\\ \nn
&+&  {\rm Re}\sum_{b}\left(\tilde{c}(-b^2/4)\frac{2U_2}{\pi}{\rm Li}_2(e^{\frac{2\pi i b V_2}{U_2}})-{c}(-b^2/4)\frac{6U_2^3}{\pi^3 Y}{\rm Li}_4(e^{\frac{2\pi i b V_2}{U_2}})\right)+c\ln[Y]+\kappa,
\end{eqnarray}
where $c,\kappa$ are constants coming from renormalization of the degenerate orbit integration as $\tau_2$ goes to complex infinity, $Y=T_2U_2-V_2^2$.
The term corresponding to $\zeta(3)$ gives the value of Euler character of the Calabi Yau dual. This was also evaluated for several standard embeddings in \cite{Chattopadhyaya:2017zul}, and non-standard embeddings in \cite{Banlaki:2018pcc} and in \cite{Banlaki:2019bxr}.

The relevant piece for computing the perturbative polynomial in the pre-potential for $K3\times T^2$ would be given by,
\begin{eqnarray}
{\cal I}_{poly}&=&{\cal I}_0+\tilde{c}(0)\frac{\pi}{3}U_2-c(0)\frac{\pi}{15} \frac{U_2^3}{Y}\\ \nn
&+& 2 {\rm Re}\sum_{b}\left(\tilde{c}(-b^2)\frac{2U_2}{\pi}{\rm Li}_2(e^{\frac{2\pi i b V_2}{U_2}})-{c}(-b^2)\frac{6U_2^3}{\pi^3 Y}{\rm Li}_4(e^{\frac{2\pi i b V_2}{U_2}})\right)\\ \nn
&=& \frac{\pi Y}{6 U_2}E_2^2 F(q)|_{q^0}+\tilde{c}(0)\frac{\pi}{3}U_2-c(0)\frac{\pi}{15} \frac{U_2^3}{Y}\\ \nn
&+&  {\rm Re}\sum_{b}\left(\tilde{c}(-b^2/4)\frac{2U_2}{\pi}{\rm Li}_2(e^{\frac{2\pi i b V_2}{U_2}})-{c}(-b^2/4)\frac{6U_2^3}{\pi^3 Y}{\rm Li}_4(e^{\frac{2\pi i b V_2}{U_2}})\right)\\ \nn
&=&\frac{1}{2Y}\left(\frac{\pi Y^2}{3 U_2}E_2^2 F(q)|_{q^0}+\tilde{c}(0)\frac{2Y\pi}{3}U_2-c(0)\frac{\pi}{15} 2U_2^3 \right.\\ \nn
&+&\left.  {\rm Re}\sum_{b}\left(\tilde{c}(-b^2/4)\frac{4YU_2}{\pi}{\rm Li}_2(e^{\frac{2\pi i b V_2}{U_2}})-{c}(-b^2/4)\frac{12U_2^3}{\pi^3 }{\rm Li}_4(e^{\frac{2\pi i b V_2}{U_2}})\right)\right) \\ \nn
&=& -\frac{1}{2Y}d_{abc}y_2^ay_2^by_2^c
\end{eqnarray}
where $y_2^i \in \{T_2,U_2, V_2\}$.
We consider $V_2<0, \frac{|V_2|}{U_2}<1$ then $0\le \theta<2\pi$ which is the desired limit of validity of these expressions.

\noindent
We have the following $q$ expansions:
\begin{eqnarray}
 \frac{E_2^2}{24 \eta^{24}}\left((12-n)E_{4,1}^{even}E_6+(12+n)E_4 E_{6,1}^{even}\right)=\frac{1}{q}+6 (-43 + 2 n)+ O(q) \\ \nn
\frac{E_2}{24 \eta^{24}}\left((12-n)E_{4,1}^{even}E_6+(12+n)E_4 E_{6,1}^{even}\right)=\frac{1}{q}+6 (-39 + 2 n)+ O(q) \\ \nn
\frac{1}{24 \eta^{24}}\left((12-n)E_{4,1}^{even}E_6+(12+n)E_4 E_{6,1}^{even}\right)=\frac{1}{q}+6 (-35 + 2 n)+ O(q) \\ \nn
\frac{E_2^2}{24 \eta^{24}}\left((12-n)E_{4,1}^{odd}E_6+(12+n)E_4 E_{6,1}^{odd}\right)=-\frac{2(8+3n)}{q^{1/4}}+ O(q^{3/4}) \\ \nn
\frac{E_2}{24 \eta^{24}}\left((12-n)E_{4,1}^{odd}E_6+(12+n)E_4 E_{6,1}^{odd}\right)=-\frac{2(8+3n)}{q^{1/4}}+ O(q^{3/4}) \\ \nn
\frac{1}{24 \eta^{24}}\left((12-n)E_{4,1}^{odd}E_6+(12+n)E_4 E_{6,1}^{odd}\right)=-\frac{2(8+3n)}{q^{1/4}}+ O(q^{3/4}) \\ \nn
\end{eqnarray}

Hence we have
\begin{eqnarray}\nn
c(0)&=&6(-35+2n)\\ \nn
 \tilde{c}(0)&=&6(-39+2n)\\ \nn
c(-1/4)&=&\tilde{c}(-1/4)=-2(8+3n)\\
c(-1)&=&\tilde{c}(-1)=1 \\ \nn
E_2^2F(q)|_{q^0}&=&-4(8+3n)+2+6(-43+2n)
\end{eqnarray}
The polynomial term from genus 1 computation can be given as:
\begin{eqnarray}\label{dabc}
\frac{1}{\pi} d_{abc}y^ay^by^c&=& 96 T_2^2 U_2 + 176 T_2 U_2^2 - 32 U_2^3 - (112+48n) T_2 U_2 V_2 \\ \nn
& -& (96-48n) T_2 V_2^2  -( 80-48n) U_2 V_2^2  - (16+48n) V_2^3
\end{eqnarray}

If $V_2$ is taken to be negative we have
\begin{eqnarray}\label{dabcn}
\frac{1}{\pi} d_{abc}y^ay^by^c&=& 96 T_2^2 U_2 + 176 T_2 U_2^2 - 32 U_2^3 + (112+48n) T_2 U_2 V_2 \\ \nn
& -& (96-48n) T_2 V_2^2  -( 80-48n) U_2 V_2^2  + (16+48n) V_2^3
\end{eqnarray}

\subsection{Orbifolds on $K3$}
The structure of the polynomial is ensured if there is no potentially problematic term of the form $V_2^4/U_2$. This needs to vanish for any embedding under any orbifold action. The result of the degenerate orbit only appears from the ${cal Z}_{\rm new}^{(0,s)}$ sectors. These terms may appear from the analogs of ${\rm Li}_2$ and ${\rm Li}_4$ but with non-integral $j$. These are possible contributions from ${\cal Z}_{\rm new}^{(0,s)}$ sectors. However a quick residue calculation from Sommerfield Watson technique we can see that,
%Using the Sommerfield-Watson technique we can show that
\begin{eqnarray}
{\rm Re}\left(\sum_{j=1}^{\infty}\frac{e^{i\theta (j+s/N)}}{(j+s/N)^n}\right)=\pi i \left({\rm Res}\left(\frac{e^{i\theta (z+s/N)}}{(z+s/N)^n(1-e^{2\pi i z})}\right)\right)
\end{eqnarray}
where $n$ is an integer, $n>1$ and $N\ne 1$ is the order of the orbifold, $0\le \theta<2\pi$.

By computing the residues we get
\begin{eqnarray}
\left(\sum_{j=1}^{\infty}\frac{e^{i\theta (j+s/N)}}{(j+s/N)^2}\right)&=&\pi i \left({\rm Res}\left(\frac{e^{i\theta (z+s/N)}}{(z+s/N)^2(1-e^{2\pi i z})}\right)\right)\\ \nn
&=& \frac{\pi}{2}  \left(-\theta +i \theta  \cot \left(\frac{\pi  s}{N}\right)+\pi  \csc ^2\left(\frac{\pi  s}{N}\right)\right)
\end{eqnarray}
and 
\begin{eqnarray}
\left(\sum_{j=1}^{\infty}\frac{e^{i\theta (j+s/N)}}{(j+s/N)^2}\right)&=&\pi i \left({\rm Res}\left(\frac{e^{i\theta (z+s/N)}}{(z+s/N)^4(1-e^{2\pi i z})}\right)\right)\\ \nn
&=& \frac{\pi  e^{\frac{2 i \pi  s}{N}}}{6 \left(-1+e^{\frac{2 i \pi  s}{N}}\right)^4}  \left((2 \pi -\theta )^3+\theta ^3 e^{\frac{6 i \pi  s}{N}}+\left(3 \theta ^3-12 \pi  \theta ^2+32 \pi ^3\right) e^{\frac{2 i \pi  s}{N}}\right.\\ \nn
&&\left.+\left(-3 \theta ^3+6 \pi  \theta ^2+12 \pi ^2 \theta +8 \pi ^3\right) e^{\frac{4 i \pi  s}{N}}\right)
\end{eqnarray}

The above residue calculations show that the problematic term $\frac{V_2^4}{U_2}$ do not appear in the calculation for orbifolds of $K3$ with some order $N$.
Hence the polynomial term always exists for any of the orbifolds of $K3$ under $g'\in[M_{23}]$.

The only question now remains is whether these coefficients $\frac{d_{abc}}{\pi}$ are rational as in equation (\ref{dabcn}).
This would also hold for any $K3$ orbifold if we can show that the following equations 
\begin{eqnarray}
\sum_{s=1}^{N-1}c^{(0,s)}(-b^2/4)\sum_{j=1}^{\infty}\frac{\cos(\theta (j+s/N))}{(j+s/N)^4}\\ \nn
\sum_{s=1}^{N-1}c^{(0,s)}(-b^2/4)\sum_{j=1}^{\infty}\frac{\cos(\theta (j+s/N))}{(j+s/N)^2}
\end{eqnarray}
have rational coefficients of $\pi$ and $\pi^3$ respectively. For the models which we studied for both standard embeddings with orbifolds of $K3$ with an action of $g'$ which corresponds to a conjugacy class of $[M_{23}]$ as well as non-standard ones for order 2 and order 3 actions this holds true.

\begin{table}[H]
	\renewcommand{\arraystretch}{0.5}
	\begin{center}
		\vspace{0.2cm}
		\begin{tabular}{|c|c|c|}
			\hline
			Order $N$ & $\sum'{\rm Re}\left(\sum_{j=1}^{\infty}\frac{e^{i\theta (j+s/N)}}{(j+s/N)^2}\right)$ &  $\sum'{\rm Re}\left(\sum_{j=1}^{\infty}\frac{e^{i\theta (j+s/N)}}{(j+s/N)^4}\right)$  \\ \hline
			& &  \\
			2 & $\frac{\pi}{2}  (\pi -\theta )$ & $\frac{1}{12} \pi  \left(\theta ^3-3 \pi  \theta ^2+2 \pi ^3\right)$\\ \nn
		
			3 & $\frac{1}{3} \pi  (4 \pi -3 \theta )$ & $\frac{1}{18} \pi  \left(3 \theta ^3-12 \pi  \theta ^2+16 \pi ^3\right)$ \\ 
							4 & ${\pi}  (2\pi -\theta )$ & $\frac{1}{6} \pi  \left(\theta ^3-6 \pi  \theta ^2+16 \pi ^3\right)$\\ \nn
			5 & $ 2\pi  (2 \pi -\theta )$ & $\frac{1}{15} \pi  \left(5 \theta ^3-30 \pi  \theta ^2+104 \pi ^3\right)$\\
				6 & $ \pi  (4 \pi - \theta )$ & $\frac{1}{3} \pi  \left(\theta ^3-12 \pi  \theta ^2+80 \pi ^3\right)$ \\ 
			7 & $\pi(8\pi-3\theta)$ & $\frac{\pi  \theta ^3}{2}+4 \pi ^2 \theta ^2+\frac{80 \pi ^4}{3}$ \\
				8 & $ 2\pi  (4 \pi -\theta )$ & $\frac{1}{3} \pi  \left(\theta ^3-12 \pi  \theta ^2+128 \pi ^3\right)$\\
			11 & $5 \pi (4 \pi - \theta)$ & $\frac{5 \pi  \theta ^3}{6}+10 \pi ^2 \theta ^2+\frac{488 \pi ^4}{3}$\\
			14 &  $24\pi ^2-3 \pi \theta$ & $-12 \pi ^2 \theta ^2+\frac{\pi}{2} \theta ^3+400 \pi ^4$ \\
			15 & $32\pi^2-4\theta\pi$ &  $\frac{2 \pi  \theta ^3}{3}-16 \pi ^2 \theta ^2+\frac{1664 \pi ^3}{3}$\\
			23 & $11 \pi  (8 \pi -\theta )$ & $\frac{9328}{3}\pi^4-44\theta^2\pi^2+\pi \frac{11}{6}\theta^3$ \\
		& & \\	
			\hline
		\end{tabular}
	\end{center}
	\vspace{0.2cm}
	\caption{ Residues corresponding to orbifolds of $K3$. $\sum'$ corresponds to summing over those sectors $s$ where ${\rm gcd}(N,s)=1$.}
	\label{residuep}
	\renewcommand{\arraystretch}{0.5}
\end{table}

For an order $N$ orbifold of $K3$ the polynomial in terms of $T_2,U_2,V_2$ is given by
\be
{\cal I}^N_{poly}=\frac{1}{N}{\cal I}_{poly}+\sum_{s=1}^{N-1}{\cal I}^{(0,s)}_{poly},
\ee
where we have,
\begin{eqnarray}\label{i0striple}
\sum_{s=1}^{N-1}{\cal I}^{(0,s)}_{poly}&=&\frac{1}{2Y}\sum_{s=1}^{N-1}\left(\tilde{c}^{(0,s)}(0)\frac{2Y\, 2 \zeta(2,s/N)}{\pi}U_2-c^{(0,s)}(0)\frac{6 \zeta(4,s/N)}{\pi^3} 2U_2^3 \right.\\ \nn
&+&\left.  {\rm Re}\sum_{b}\left(\tilde{c}^{(0,s)}(-b^2)\frac{4YU_2}{\pi}{\rm Li}^s_2(e^{\frac{2\pi i b V_2}{U_2}})-{c}^{(0,s)}(-b^2)\frac{12U_2^3}{\pi^3 }{\rm Li}^s_4(e^{\frac{2\pi i b V_2}{U_2}})\right)\right) ,
\end{eqnarray}
with
\be
{\rm Re}\left({\rm Li}^s_k(e^{i\theta})\right)=\sum_{j=1}^{\infty}\frac{e^{i\theta (j+s/N)}}{(j+s/N)^k}.
\ee
For standard embeddings the values of $c^{(0,s)}(-b^2)$ and $\tilde c^{(0,s)}(-b^2)$ are listed in
 Appendix \ref{lowcf}.

\subsection{Non-standard embeddings for orbifolds of order 2 and 3}
Several non-standard embeddings of order $2$ and $3$ orbifolds of $K3$ were studied in \cite{Chattopadhyaya:2016xpa,Banlaki:2019bxr} respectively. The twisted elliptic genus of these models which feeds in the standard embeddings are given by the 2A and 3A conjugacy classes of $[M_{23}]$. In presence of a Wilson line the result of ${\cal Z}_{\rm new}^{(r,s)}$ and their perturbative polynomial would be given in this section.

\vspace{0.3cm}
\noindent
{\bf 2A orbifold:}
For the $O(2)$ $g'$ orbifold of $K3$ the new supersymmetric index was given by,
\begin{eqnarray} \label{abc}
& & {\cal Z}_{\rm new} 
=  \frac{1}{ 2\eta^{24}} \left\{ \Gamma^{(0,0)}_{3, 2}  \otimes \frac{1}{12}
[(12-n) \hat E_{4, 1}  E_6 +(12+n) \hat E_{6, 1} E_{4}] 
\right. \\ \nonumber
& &   + \Gamma^{(0,1)}_{3, 2} \otimes  \left[
\hat a \hat E_{4,1}(E_6+2{\cal E}_2(\tau) E_4)+
\hat b {\cal E}_2(\tau)^2 (\hat E_{6,1}+2 {\cal E}_2(\tau) \hat E_{4,1})+
\hat c E_{4}(\hat E_{6,1}+2 {\cal E}_2(\tau) \hat E_{4,1}) \right] \\ \nn
&& + {\rm Modular\; transformations}
\end{eqnarray}
The parameters  $\hat a , \hat c$  are given by 
\begin{eqnarray} \label{eqnac}
\hat a=\frac{12-n}{36}- \frac{\hat b}{2}, \qquad
\hat c=  \frac{2}{3} - \hat a - \hat b \;.
\end{eqnarray}

Its low order Fourier coefficients are given by
\begin{eqnarray}
c^{(0,1)}(-1)=\tilde{c}^{(0,1)}(-1)=1\\ \nn
c^{(0,1)}(-1/4)=\tilde{c}^{(0,1)}(-1/4)=-2 (-16 + 18 \hat b + n) \\ \nn
c^{(0,1)}(0)=206 - 216 \hat b + 4 n\\ \nn
\tilde{c}^{(0,1)}(0)=182 - 216 \hat b + 4 n
\end{eqnarray}
where $\hat b$ corresponds to the type of the shift and $n$ the instanton embedding as in \cite{Chattopadhyaya:2016xpa} listed in table...

Using these results we have for $2A$ orbifold
\begin{eqnarray}
{\cal I}^{(0,1)}_{poly}&=& \frac{ \pi}{Y}   \left(T_2 U_2 \left(8 V_2 (18 b+n-17)+(247-288 b) U_2\right)\right.\\ \nn
&&\left.+U_2 V_2^2 (-144 b-24 n+185)+8 V_2^3 (18 b+n-23)+16 (18 b-17) U_2^3\right)
\end{eqnarray}
\be
{\cal I}^{2A}_{poly}=\frac{1}{2}{\cal I}_{poly}+{\cal I}^{(0,1)}_{poly}=-\frac{1}{2Y}d^{2A}_{abc}y_2^a y_2^b y_2^c
\ee

\vspace{0.3cm}
\noindent
{\bf 3A orbifold:} We take the model studied in \cite{Chattopadhyaya:2017zul} where $K3$ is taken as an orbifold limit of $T^4/\mathbb{Z}_3$ of the 4 bosons in the heterotic compactification. The computation of new supersymmetric index follows the same method as in \cite{Chattopadhyaya:2017zul} with some modifications due to the presence of Wilson lines and will be briefly discussed in Appendix \ref{znew3}.
In presence of a single Wilson line the new super-symmetric index in the untwisted sector for 3A orbifold on $K3$ are given by:
\begin{eqnarray}\nn
{\cal Z}_{\rm new}^{(0,1)}&=&\frac{1}{\eta^{24}}\left(\hat a \hat E_{4,1}E_6+\hat b E_{6,1}E_4+\hat c \hat E_{4,1}E_4(3\tau){\cal E}_3(\tau)+\hat d E_6(3\tau)\hat E_{4,1}+\hat f E_4(3\tau)\hat E_{6,1}  \right)\\
\end{eqnarray}
where different values of $\hat a,\hat b,\hat c,\hat d, \hat f$ are given in the following table:
\begin{table}[H]
	\renewcommand{\arraystretch}{0.5}
	\begin{center}
		\vspace{0.5cm}
		\begin{tabular}{|c|c|c|ccccc|}
			\hline
			& & & & & & & \\
			Shift & $N_h-N_v$ & $n$ & $\hat a$ & $\hat b$ & $\hat c$ & $\hat d$ & $\hat f$\\ \hline
			& & & & & & & \\
			(1,-1,$0^6)\times (0^8$)& $-134$ & $-12$ & $\frac{1}{28}$ & 0 & $\frac{27}{4}$ & $-\frac{81}{14}$  & 0\\
			& & & & & & & \\
				$(2,1,1,1,1,0^3)\times (0^8)$& $-80$ & $-12$ & $\frac{-3}{80}$ & 0 & $-\frac{81}{40}$ & $-\frac{243}{80}$  & 0\\
			& & & & & & &  \\
			(2,$0^7)\times(2,0^7$)& 64  & 0 & $-\frac{57}{1680}$ &$\frac{3}{80}$  &$\frac{63}{40}$ &$-\frac{513}{560}$ &$\frac{27}{80}$  \\
			& & & & & & & \\
			(1,-1,$0^6)\times(2,1,1,0^5$)& 28 & 6& $-\frac{137}{1120}$ &$\frac{39}{320}$ &$\frac{27}{40}$ &$\frac{81}{1120}$  & $\frac{81}{320}$\\
			& & & & & & & \\
			$(2,1,1,0^5)\times(2,1,1,1,1,0^3)$& 82 &3 &$\frac{-117}{2240}$ &$\frac{37}{640}$ &$\frac{81}{40}$ &$\frac{-3159}{2240}$ & $\frac{243}{640}$ \\
			& & & & & & & \\
			\hline
		\end{tabular}
	\end{center}
	\vspace{-0.5cm}
	\caption{${\cal Z}^{r,s}_{\rm new}$ for different embeddings with $K3$ as $T^4/Z_3$ and $N=3$ CHL orbifold with one Wilson line}\label{znew3wil}
	\renewcommand{\arraystretch}{0.5}
\end{table}

Hence the low lying Fourier coefficients are given as
\begin{eqnarray}
c^{(0,1)}(-1)=\tilde{c}^{(0,1)}(-1)=1\\ \nn
c^{(0,1)}(0)= -6 (59 \hat a + 11 \hat b - 27 \hat c - 25 \hat d + 51 \hat f) \\ \nn
\tilde{c}^{(0,1)}(0)=-6 (63 \hat a + 15 \hat b - 23 \hat c - 21 \hat d + 55 \hat f) \\ \nn
c^{(0,1)}(-1/4)=\tilde{c}^{(0,1)}(-1/4)=8 (7 \hat a - 11 \hat b + 7 (\hat c +\hat d) - 11 \hat f).\\ \nn
\end{eqnarray}
Hence the polynomial term in the gravitational coupling correction at $g=1$ is given by,
\begin{eqnarray}
{\cal I}_{poly}^{3A}=\frac{1}{3}{\cal I}_{poly}+\sum_{s=1}^{2}{\cal I}_{poly}^{0,s}=-\frac{1}{2Y}d^{3A}_{abc}y_2^a y_2^b y_2^c.
\end{eqnarray}
In the above equation we have,
\begin{eqnarray}
\sum_{s=1}^2{\cal I}_{poly}^{(0,s)}&=& \frac{16}{3Y} \pi  \left(T_2 U_2 \left(U_2 (-133 \hat a-133 \hat b+125 \hat c+119 \hat d-253 \hat f+1)\right.\right.\\ \nn
&&\left.\left.-3 V_2 (28 \hat a-44 \hat b+28 \hat c+28 \hat d-44\hat f+1)\right)\right. \\ \nn
&&\left.+U_2 V_2^2 (469 \hat a-395 \hat b+211 \hat c+217 \hat d-275 \hat f+23)\right.\\ \nn
&&\left.+U_2^3 (242 \hat a+242\hat b-274 \hat c-262\hat d+482\hat f-2)\right.\\ \nn
&&\left.-3 V_2^3 (28 \hat a-44\hat b+28\hat c+28\hat d-44\hat f+7)\right)
\end{eqnarray}
with
$\hat a+\hat b+\hat c+\hat d+\hat f=1$.

\section{Conclusion}\label{con}
The twisted elliptic genus of $K3$ under different orbifolds corresponding to 26 conjugacy classes of Mathieu group $M_{24}$ are of interest to both physics and mathematics community for its rich modular and combinatoric structures. It is evident that these coefficients are of extreme importance even for the heterotic-type II duality symmetries for standard embeddings. The new super-symmetric index and twisted elliptic genus are related to each other with spectral flow and characters of $D_6$ lattice in various sectors as well as broken $E_8$ partition functions. Now since the predicted Gopakumar Vafa invariants of the dual Calabi Yau geometry can be obtained from the new super-symmetric index it may be interesting to study how these characters of Mathieu group is involved in counting holomorphic curves in the dual Calabi Yau picture.

In this work we studied the Wilson line addition to $E_8$ gauge group in different chains and at each step an additional Wilson line being added for standard embeddings. It may be interesting to explore the role of the twisted elliptic genus for non-standard embeddings too when Wilson lines are added. It is however computationally challenging to extract the Gopakumar Vafa invariants from the Calabi Yau geometry with standard embedding than with the non-standard ones where both the $E_8$ gauge groups are broken. However from the heterotic side we have shown that the integrality of the Gopakumar-Vafa invariants are ensured as also the triple intersection polynomial can be computed from the heterotic side. However there remain subtleties in identifying the exact triple intersection numbers due to several conifold singularities where the transition from one Weyl chamber to another is expected. For a STU model in $K3\times T^2$. This transition is only observed at $T=U$.
 Here however there can be singularities at $\frac{r}{N}T=U$ for each $r=1,...N-1$ in general and there is no singularity at $T=U$ for $g'\in[M_{23}]$. At these singular points transitions are possible from one Weyl chamber to another and the perturbative pre-potential will also pick up terms linear or quadratic in moduli $T,U,V$ and $TU-V^2$ \cite{Harvey-Moore}. Our result for $p_n(T,U,V)$ is valid at the limit $T_2/N>U_2$ and $|V_2|<<U_2$
where the degenerate orbit of one loop computation of $F_1$ contains a polynomial of degree 3 which indicates the presence of a dual Calabi Yau geometry.
We hope to explore more on the monodromies for these orbifold models in future.

Another interesting direction of exploration could be the action of these orbifolds for gauge couplings. In our previous work \cite{Chattopadhyaya:2016xpa} the difference of one loop gauge threshold corrections for standard and $N=2 $ non-standard embeddings were studied. In the recent work of \cite{Angelantonj:2016gkz} the gauge corrections with non-zero 3-form flux were computed. We hope to generalize this to $N=3$ non-standard embeddings in future.

\acknowledgments

The author thanks Prof. Justin David, Jan Manschot and Thorsten Schimannek for helpful discussions. The author also thanks Apratim Kaviraj for help with some Mathematica codes and diagrams used in this paper. The work of the author is funded by Laureate Award 15175 of the Irish Research Council.

\appendix

\section{New supersymmetric index for 3A orbifold of $K3$}\label{znew3}

We consider an orbifold limit of $K3$ and the 6 compact directions other than the two $E_8$ lattices are $\left[(T^4/\mathbb{Z}_3)\times T^2\right]/\mathbb{Z}_3$.
The orbifold action on the coordinates of the torus can be given by
\be \label{eqn:tor_act}
g:\,(z_1,\,z_2,\,z_3)\mapsto(e^{2\pi i/3}z_1,\,e^{-2\pi i/3}z_2,\,x_3,\,x_4)\,,
\ee
where $x_3,x_4$ are coordinates on $T^2$ and complex bosons $z_1,z_2$ parametrize $T^4$.
The cycles of $T^4$ form an $SU(3)$ lattice in $\mathbb{C}^2$ which is given by
\begin{align}
e_1=e^{\frac{2\pi i}{3}}\,,\quad e_2=1\,.
\end{align}
The combined action denotes a $1/3$ shift in one of the circles of $T^2$ and $g'$ on $K3$ which is as follows:
\be \label{eqn:chl_act}
\left(z_1+\frac{1}{3}e_1+\frac{2}{3}e_2,\,z_2,\,x_3+\frac{1}{3},\, x_4\right)\,.
\ee

The new supersymmetric index without Wilson line in the order 3 orbifold of $K3$ was given for non-standard embeddings in \cite{Banlaki:2019bxr}. For $s\ne 0$ we have
\begin{eqnarray}\nn
{\cal Z}_{\rm new}^{0,s}&=&\frac{4}{3}\left(\hat a E_4E_6+\hat b {\cal E}_3^2(\tau)E_6+\hat c E_4(E_6+3{\cal E}_3(\tau)E_4)+\hat d {\cal E}_3^2(\tau)(E_6+3{\cal E}_3(\tau)E_4)\right)\\
\end{eqnarray}
In presence of Wilson lines this can be computed as using the formula
\begin{eqnarray} \nn 
{\cal Z}_{\rm new}(q, \bar q)  = -\frac{1}{2\eta^{20}(\tau)}\sum_{a,b=0}^2 \sum_{r,s=0}^2 e^{-\frac{2\pi iab}{9}} 
Z_{E_8}^{(a,b)} (\tau)  Z_{E_8'}^{(a,b)} (\tau) 
\times \frac{1}{9} F(a,r,b,s,q)\otimes \Gamma^{(r,s)}_{3,2} (q,\bar q),\\ \label{z3nswilson}
\end{eqnarray}
Different components of the partition function are given by,
\begin{eqnarray} \label{fabrshet}
F ( a, r, b, s; q)  =
k^{(a,r,b,s)}\eta^2(\tau)q^{\frac{-a^2}{9}}\frac{1}{\theta_1 ^2(\frac{a\tau+b}{3}, \tau)}.
\end{eqnarray}

The partition function over the shifted $E_8$ lattices are given by
\begin{eqnarray}
Z_{E_8}^{a,b}(q)&=&\frac{1}{3}\sum_{\al,\beta=0}^{1}
e^{-i\pi\beta\frac{a}{3}\sum_{I=1}^8\gamma^{I}}
\prod_{I=1}^8\theta\left[
\begin{smallmatrix}\al+2\frac{a}{\nu}\gamma^I \\ \beta+2\frac{b}{3}\gamma^I\end{smallmatrix}\right],\\ 
Z_{E_8'}^{a,b}(q)&=&\frac{1}{3}\sum_{\al,\beta=0}^{1}e^{-i\pi\beta\frac{a}{3}
	\sum_{I=1}^8\tilde{\gamma}^{I}}\prod_{I=1}^6\theta\left[
\begin{smallmatrix}\al+2\frac{a}{\nu}\tilde{\gamma}^I \\ \beta+2\frac{b}{3}\tilde{\gamma}^I 
\end{smallmatrix}\right]\theta^2\left[
\begin{smallmatrix}\al \\ \beta
\end{smallmatrix};y\right],
\end{eqnarray}
where $\gamma, \tilde\gamma$ are the shifts in the two $E_8$ lattices.

The phases $k^{a,b,r,s}$ are important for modular transformation among different sectors such that the whole partition function remain modular invariant.
Under non-standard embeddings these are given by,
\begin{eqnarray}
k^{a,0,b,s}&=&9\left(\begin{matrix}
0 & 1 & 1 \\
1 & e^{-\pi i(2-\Gamma^2)/9} & e^{-2\pi i(2-\Gamma^2)/9} \\
1 & e^{-5\pi i(2-\Gamma^2)/9} & e^{-4\pi i(2-\Gamma^2)/9}
\end{matrix}\right)\\ \nn
k^{a,0,b,1}&=& k^{a,0,b,2}=9\left(\begin{matrix}
0 & 1 & 1 \\
0 & 0 & 0\\
0 & 0 & 0
\end{matrix}\right)\\ \nn
k^{a,1,b,0}&=& k^{a,2,b,0}=9\left(\begin{matrix}
0 & 0 & 0 \\
1 & 0 & 0\\
1 & 0 & 0
\end{matrix}\right)\\ \nn
k^{a,1,b,1}&=&k^{a,2,b,1}=9\left(\begin{matrix}
0 & 0 & 0 \\
0 & e^{-\pi i(2-\Gamma^2)/9} & 0\\ 
0 & 0 & e^{-4\pi i(2-\Gamma^2)/9}
\end{matrix}\right)\\ \nn
k^{a,1,b,2}&=&k^{a,2,b,1}=9\left(\begin{matrix}
0 & 0 & 0 \\
0 & 0 & e^{-2\pi i(2-\Gamma^2)/9} \\ 
0 & e^{-5\pi i(2-\Gamma^2)/9} & 0
\end{matrix}\right)
\end{eqnarray}
where $\Gamma^2=\gamma^2+\tilde{\gamma^2}$ are the square of all the shifts in the two $E_8$ lattices.

The lattice involving the winding and momenta modes are given in each case by,
\begin{eqnarray} \label{gamma32}
\Gamma_{3,2}^{(r,s)} (q, \bar q) 
&=& \sum_{\stackrel{m_1, m_2, n_2 \in \mathbb{Z} }{n_1= \mathbb{Z} + \frac{r}{3} } }
q^\frac{p_L^2}{2} \bar q ^{\frac{p_R^2}{2}} e^{2\pi i m_1 s /3}, 
\\ \nn
\frac{1}{2} p_R^2 &=& 
\frac{1}{2(T_2 U_2-V_2^2)} |-m_1 U + m_2 + n_1 T + n_2 (TU-V^2)+b V |^2 , \\ \label{plpr}
\frac{1}{2}p_L^2 &=& \frac{1}{2} p_R^2 + m_1n_1 + m_2 n_2+\frac{b^2}{4}
\end{eqnarray}
and $T, U$ being Kahler and complex structure of $T^2$ and $V$ is the Wilson line moduli.

\section{List of genus zero GV invariants:}\label{gv}
We shall remove the overall negative sign from each of the $n^0_{n_1,n_2,b}$ in the following tables.
\subsection{Standard embeddings (2A,3A)}
\begin{table}[H]
	\renewcommand{\arraystretch}{0.5}
	\begin{center}
		\vspace{0.5cm}
		\setlength{\doublerulesep}{20\arrayrulewidth}
		{\scriptsize{
				\begin{tabular}{|c|c|c|c|c|c|}
					\hline
					& & & & & \\
					$(n_1,n_2,b)$ & $(0,0,0)$& $(1,1,0)$ & $(2,1,0)$ & $(1,2,0)$ & $(2,2,0)$  \\
					\hline
					& & & & & \\
					$n^0_{(n_1,n_2,b)}$ & $-196$ & 93184 &4683776 & 4560128 & 2011496600 \\
					%			& & &   \\
					\hline
					\hline
					& & & & & \\
					$(n_1,n_2,b)$ & $(1/2,0,0)$ & $(1/2,2,0)$ & $(1/2,4,0)$ & $(1/2,6,0)$ & $(1/2,8,0)$\\
					\hline
					& & & & & \\
					$n^0_{(n_1,n_2,b)}$ & 512 & 93184 & 4683776 & 119394304 & 2018028544  \\
					\hline
					\hline
					& & & & & \\
					$(n_1,n_2,b)$ & $(1/2,-1,0)$ & $(1/2,1,0)$ & $(3/2,1,0)$ & $(5/2,1,0)$ & $(7/2,1,0)$ \\
					\hline
					& & & & & \\
					$n^0_{(n_1,n_2,b)}$ & 16 & 6304 & 701280 & 24821184  & 508989392  \\
					\hline
					\hline
					& & & & & \\
					$(n_1,n_2,b)$ & $(0,0,1)$& $(1,1,1)$ & $(2,1,1)$ & $(1,2,1)$ & $(2,2,1)$  \\
					\hline
					& & & & & \\
					$n^0_{(n_1,n_2,b)}$ & $-112$ & 28672 & 1900544 & 1830640 & 1021517712 \\
					%			& & &   \\
					\hline
					\hline
					& & & & & \\
					$(n_1,n_2,b)$ & $(1/2,0,1)$ & $(1/2,2,1)$ & $(1/2,4,1)$ & $(1/2,6,1)$ & $(1/2,8,1)$\\
					\hline
					& & & & & \\
					$n^0_{(n_1,n_2,b)}$ & 0 & 28672 & 1900544 & 55595008 & 1025703936   \\
					\hline
					\hline
					& & & & & \\
					$(n_1,n_2,b)$ & $(1/2,-1,1)$ & $(1/2,1,1)$ & $(3/2,1,1)$ & $(5/2,1,1)$ & $(7/2,1,1)$ \\
					\hline
					& & & & & \\
					$n^0_{(n_1,n_2,b)}$ & 0 & 896 & 249344  & 10832256  & 248294912    \\
					\hline
				\end{tabular}
		}}
		
	\end{center}
	\vspace{-0.2cm}
	\caption{$2A$ orbifold, standard embedding, with single Wilson line.}
	\renewcommand{\arraystretch}{0.5}
\end{table}

\begin{table}[H]
	\renewcommand{\arraystretch}{0.5}
	\begin{center}
		\vspace{0.5cm}
		\setlength{\doublerulesep}{20\arrayrulewidth}
		{\scriptsize{
				\begin{tabular}{|c|c|c|c|c|c|}
					
					\hline
					& & & & & \\
					$(n_1,n_2,b)$ & $(0,0,0)$ & $(1,1,0)$ & $(1,2,0)$ & $(1,3,0)$ & $(1,4,0)$ \\
					\hline
					& & & & & \\
					$n^0_{(n_1,n_2,b)}$ & $-48$ & 67554 & 3261546 &  74765388 & 1363824864   \\
					\hline
					\hline
					& & & & & \\
					$(n_1,n_2,b)$ & $(r/3,0,0)$ & (1/3,3,0) & (2/3,3,0) & (1/3,9,0) & (2/3,6,0) \\
					\hline
					& & & & & \\
					$n^0_{(n_1,n_2,b)}$ & 378 & 67554  & 3261546 & 81503442 & 1363824864 \\
					\hline
					\hline
					& & & & & \\
					$(n_1,n_2,b)$ & $(1/3,-1,0)$ & $(2/3,-1,0)$ & $(1/3,1,0)$ & (2/3,1,0) & (4/3,1,0)  \\
					\hline
					& & & & & \\
					$n^0_{(n_1,n_2,b)}$ & 18 & $6$ & 2382 & 7056 & 270012  \\
					\hline
					& & & & & \\
					$(n_1,n_2,b)$ & $(0,0,1)$ & $(1,1,1)$ & $(1,2,1)$ & $(1,3,1)$ & $(1,4,1)$ \\
					\hline
					& & & & & \\
					$n^0_{(n_1,n_2,b)}$ & $-48$ & 67554 & 3261546 &  74765388 & 1363824864   \\
					\hline
					\hline
					& & & & & \\
					$(n_1,n_2,b)$ & $(r/3,0,1)$ & (1/3,3,1) & (2/3,3,1) & (1/3,9,1) & (2/3,6,1) \\
					\hline
					& & & & & \\
					$n^0_{(n_1,n_2,b)}$ & 0 & 21168  & 1333584 & 38094624 & 694536768  \\
					\hline
					\hline
					& & & & & \\
					$(n_1,n_2,b)$ & $(1/3,-1,1)$ & $(2/3,-1,1)$ & $(1/3,1,1)$ & (2/3,1,1) & (4/3,1,1)  \\
					\hline
					& & & & & \\
					$n^0_{(n_1,n_2,b)}$ & 0 & $0$ & 336 & 1008 & 94512  \\
					\hline
					\hline
				\end{tabular}
		}}
	\end{center}
	\vspace{-0.5cm}
	\caption{$3A$ orbifold, standard embedding with single Wilson line. }
	\renewcommand{\arraystretch}{0.5}
\end{table}

\subsection{Non-standard embeddings for orbifolds of order $2$}

\begin{table}[H]
	\renewcommand{\arraystretch}{0.5}
	\begin{center}
		\vspace{0.5cm}
		\setlength{\doublerulesep}{20\arrayrulewidth}
		{\scriptsize{
				\begin{tabular}{|c|c|c|c|c|c|}
					\hline
					& & & & & \\
					$(n_1,n_2,b)$ & $(0,0,0)$& $(1,1,0)$ & $(2,1,0)$ & $(1,2,0)$ & $(2,2,0)$  \\
					\hline
					& & & & & \\
					$n^0_{(n_1,n_2,b)}$ &  4 & 92608 & 4681088 & 4554752 & 2011430936 \\
					%			& & &   \\
					\hline
					\hline
					& & & & & \\
					$(n_1,n_2,b)$ & $(1/2,0,0)$ & $(1/2,2,0)$ & $(1/2,-1,0)$ & $(1/2,1,0)$ & $(3/2,1,0)$\\
					\hline
					& & & & & \\
					$n^0_{(n_1,n_2,b)}$ &  416 & 92608 & 16 & 6304 &701280   \\
					\hline
					\hline
					& & & & & \\
					$(n_1,n_2,b)$ & $(0,0,1)$& $(1,1,1)$ & $(2,1,1)$ & $(1,2,1)$ & $(2,2,1)$  \\
					\hline
					& & & & & \\
					$n^0_{(n_1,n_2,b)}$ &  $-16$ & 29056 &  1902416 & 1834384 & 1021566960  \\
					%			& & &   \\
					\hline
					\hline
					& & & & & \\
					$(n_1,n_2,b)$ &  $(1/2,0,1)$ & $(1/2,2,1)$ & $(1/2,-1,1)$ & $(1/2,1,1)$ & $(3/2,1,1)$   \\
					\hline
					& & & & & \\
					$n^0_{(n_1,n_2,b)}$ & 48 & 29056 & 0 & 896 & 249344  \\
					\hline
				\end{tabular}
		}}
		
	\end{center}
	\vspace{-0.2cm}
	\caption{$2A$ orbifold, non-standard embedding, with single Wilson line. Shift is given by, $\gamma=\{1,1,0^6\},\tilde{\gamma}=\{2,2,0^6\}$}
	\renewcommand{\arraystretch}{0.5}
\end{table}

\begin{table}[H]
	\renewcommand{\arraystretch}{0.5}
	\begin{center}
		\vspace{0.5cm}
		\setlength{\doublerulesep}{20\arrayrulewidth}
		{\scriptsize{
				\begin{tabular}{|c|c|c|c|c|c|}
					\hline
					& & & & & \\
					$(n_1,n_2,b)$ & $(0,0,0)$& $(1,1,0)$ & $(2,1,0)$ & $(1,2,0)$ & $(2,2,0)$  \\
					\hline
					& & & & & \\
					$n^0_{(n_1,n_2,b)}$ & 164 & 90176 & 4650624 & 4587904 & 201309128  \\
					\hline
					\hline
					& & & & & \\
					$(n_1,n_2,b)$ & $(1/2,0,0)$ & $(1/2,2,0)$ & $(1/2,-1,0)$ & $(1/2,1,0)$ & $(3/2,1,0)$\\
					\hline
					& & & & & \\
					$n^0_{(n_1,n_2,b)}$ & 352 & 90176 & 8 & 6736 & 710064    \\
					\hline
					\hline
					& & & & & \\
					$(n_1,n_2,b)$ & $(0,0,1)$& $(1,1,1)$ & $(2,1,1)$ & $(1,2,1)$ & $(2,2,1)$  \\
					\hline
					& & & & & \\
					$n^0_{(n_1,n_2,b)}$ & $-32$ & 27776 &  1884784 & 1850144 & 1022590944  \\
					%			& & &   \\
					\hline
					\hline
					& & & & & \\
					$(n_1,n_2,b)$ &  $(1/2,0,1)$ & $(1/2,2,1)$ & $(1/2,-1,1)$ & $(1/2,1,1)$ & $(3/2,1,1)$   \\
					\hline
					& & & & & \\
					$n^0_{(n_1,n_2,b)}$ & 16 & 27776 & 0 & 1216 & 254720   \\
					\hline
				\end{tabular}
		}}
		
	\end{center}
	\vspace{-0.2cm}
	\caption{$2A$ orbifold, non-standard embedding, with single Wilson line. Shift is given by, $\gamma=\{1,1,0^6\},\tilde{\gamma}=\{4,0^7\},$ $\gamma=\{3,1,0^6\},\tilde{\gamma}=\{4,0^7\}$}
	\renewcommand{\arraystretch}{0.5}
\end{table}

\begin{table}[H]
	\renewcommand{\arraystretch}{0.5}
	\begin{center}
		\vspace{0.5cm}
		\setlength{\doublerulesep}{20\arrayrulewidth}
		{\scriptsize{
				\begin{tabular}{|c|c|c|c|c|c|}
					\hline
					& & & & & \\
					$(n_1,n_2,b)$ & $(0,0,0)$& $(1,1,0)$ & $(2,1,0)$ & $(1,2,0)$ & $(2,2,0)$  \\
					\hline
					& & & & & \\
					$n^0_{(n_1,n_2,b)}$ & 228 & 90368 & 4651520 & 4589696 & 2013113176  \\
					\hline
					\hline
					& & & & & \\
					$(n_1,n_2,b)$ & $(1/2,0,0)$ & $(1/2,2,0)$ & $(1/2,-1,0)$ & $(1/2,1,0)$ & $(3/2,1,0)$\\
					\hline
					& & & & & \\
					$n^0_{(n_1,n_2,b)}$ &  384 & 90368 & 8 & 6736 & 710064   \\
					\hline
					\hline
					& & & & & \\
					$(n_1,n_2,b)$ & $(0,0,1)$& $(1,1,1)$ & $(2,1,1)$ & $(1,2,1)$ & $(2,2,1)$  \\
					\hline
					& & & & & \\
					$n^0_{(n_1,n_2,b)}$ &  64 & 27648 &1884160 & 1848896  & 1022574528  \\
					%			& & &   \\
					\hline
					\hline
					& & & & & \\
					$(n_1,n_2,b)$ &  $(1/2,0,1)$ & $(1/2,2,1)$ & $(1/2,-1,1)$ & $(1/2,1,1)$ & $(3/2,1,1)$   \\
					\hline
					& & & & & \\
					$n^0_{(n_1,n_2,b)}$ & 0 & 27648 & 0 & 1216 & 254720   \\
					\hline
				\end{tabular}
		}}
		
	\end{center}
	\vspace{-0.2cm}
	\caption{$2A$ orbifold, non-standard embedding, with single Wilson line. Shift is given by, $\gamma=\{3,1,0^6\},\tilde{\gamma}=\{2,2,0^6\}$ }
	\renewcommand{\arraystretch}{0.5}
\end{table}

\begin{table}[H]
	\renewcommand{\arraystretch}{0.5}
	\begin{center}
		\vspace{0.5cm}
		\setlength{\doublerulesep}{20\arrayrulewidth}
		{\scriptsize{
				\begin{tabular}{|c|c|c|c|c|c|}
					\hline
					& & & & & \\
					$(n_1,n_2,b)$ & $(0,0,0)$& $(1,1,0)$ & $(2,1,0)$ & $(1,2,0)$ & $(2,2,0)$  \\
					\hline
					& & & & & \\
					$n^0_{(n_1,n_2,b)}$ & 148 & 88672 & 4634048 & 4601792 & 2013888632   \\
					\hline
					\hline
					& & & & & \\
					$(n_1,n_2,b)$ & $(1/2,0,0)$ & $(1/2,2,0)$ & $(1/2,-1,0)$ & $(1/2,1,0)$ & $(3/2,1,0)$\\
					\hline
					& & & & & \\
					$n^0_{(n_1,n_2,b)}$ & 272 & 88672 & 4 & 6952 & 714456    \\
					\hline
					\hline
					& & & & & \\
					$(n_1,n_2,b)$ & $(0,0,1)$& $(1,1,1)$ & $(2,1,1)$ & $(1,2,1)$ & $(2,2,1)$  \\
					\hline
					& & & & & \\
					$n^0_{(n_1,n_2,b)}$ & 8 & 27328 & 1876904 & 1859896 & 1023127560  \\
					%			& & &   \\
					\hline
					\hline
					& & & & & \\
					$(n_1,n_2,b)$ &  $(1/2,0,1)$ & $(1/2,2,1)$ & $(1/2,-1,1)$ & $(1/2,1,1)$ & $(3/2,1,1)$   \\
					\hline
					& & & & & \\
					$n^0_{(n_1,n_2,b)}$ & 24 & 27328  & 0 & 1376 & 257408   \\
					\hline
				\end{tabular}
		}}
		
	\end{center}
	\vspace{-0.2cm}
	\caption{$2A$ orbifold, non-standard embedding, with single Wilson line. Shift is given by, $\gamma=\{2,1,1,0^5\},\tilde{\gamma}=\{2,0^7\}$ }
	\renewcommand{\arraystretch}{0.5}
\end{table}

\begin{table}[H]
	\renewcommand{\arraystretch}{0.5}
	\begin{center}
		\vspace{0.5cm}
		\setlength{\doublerulesep}{20\arrayrulewidth}
		{\scriptsize{
				\begin{tabular}{|c|c|c|c|c|c|}
					\hline
					& & & & & \\
					$(n_1,n_2,b)$ & $(0,0,0)$& $(1,1,0)$ & $(2,1,0)$ & $(1,2,0)$ & $(2,2,0)$  \\
					\hline
					& & & & & \\
					$n^0_{(n_1,n_2,b)}$ & 180 & 87456 & 4618816 & 4617024 & 2014702392  \\
					\hline
					\hline
					& & & & & \\
					$(n_1,n_2,b)$ & $(1/2,0,0)$ & $(1/2,2,0)$ & $(1/2,-1,0)$ & $(1/2,1,0)$ & $(3/2,1,0)$\\
					\hline
					& & & & & \\
					$n^0_{(n_1,n_2,b)}$ &  192 & 87168 & 0 & 7168 & 718848   \\
					\hline
					\hline
					& & & & & \\
					$(n_1,n_2,b)$ & $(0,0,1)$& $(1,1,1)$ & $(2,1,1)$ & $(1,2,1)$ & $(2,2,1)$  \\
					\hline
					& & & & & \\
					$n^0_{(n_1,n_2,b)}$ & 24 & 26688 & 1868088 & 1868712 & 1023651864  \\
					%			& & &   \\
					\hline
					\hline
					& & & & & \\
					$(n_1,n_2,b)$ &  $(1/2,0,1)$ & $(1/2,2,1)$ & $(1/2,-1,1)$ & $(1/2,1,1)$ & $(3/2,1,1)$   \\
					\hline
					& & & & & \\
					$n^0_{(n_1,n_2,b)}$ & 32 & 26880 & 0 & 1536 & 260096   \\
					\hline
				\end{tabular}
		}}
		
	\end{center}
	\vspace{-0.2cm}
	\caption{$2A$ orbifold, non-standard embedding, with single Wilson line. Shift is given by, $\gamma=\{2,1,1,0^5\},\tilde{\gamma}=\{2,2,2,0^5\}$ }
	\renewcommand{\arraystretch}{0.5}
\end{table}

\begin{table}[H]
	\renewcommand{\arraystretch}{0.5}
	\begin{center}
		\vspace{0.5cm}
		\setlength{\doublerulesep}{20\arrayrulewidth}
		{\scriptsize{
				\begin{tabular}{|c|c|c|c|c|c|}
					\hline
					& & & & & \\
					$(n_1,n_2,b)$ & $(0,0,0)$& $(1,1,0)$ & $(2,1,0)$ & $(1,2,0)$ & $(2,2,0)$  \\
					\hline
					& & & & & \\
					$n^0_{(n_1,n_2,b)}$  & 196 & 224 & 87360 & 4617472  &   2014707864   \\
					\hline
					\hline
					& & & & & \\
					$(n_1,n_2,b)$ & $(1/2,0,0)$ & $(1/2,2,0)$ & $(1/2,-1,0)$ & $(1/2,1,0)$ & $(3/2,1,0)$\\
					\hline
					& & & & & \\
					$n^0_{(n_1,n_2,b)}$ &  224 & 87360 & 0 & 7168 & 718848   \\
					\hline
					\hline
					& & & & & \\
					$(n_1,n_2,b)$ & $(0,0,1)$& $(1,1,1)$ & $(2,1,1)$ & $(1,2,1)$ & $(2,2,1)$  \\
					\hline
					& & & & & \\
					$n^0_{(n_1,n_2,b)}$ & 16 & 26752 & 1868400 & 1868400 & 1023647760  \\
					%			& & &   \\
					\hline
					\hline
					& & & & & \\
					$(n_1,n_2,b)$ &  $(1/2,0,1)$ & $(1/2,2,1)$ & $(1/2,-1,1)$ & $(1/2,1,1)$ & $(3/2,1,1)$   \\
					\hline
					& & & & & \\
					$n^0_{(n_1,n_2,b)}$ & 16 & 26752 & 0 & 1536 & 260096    \\
					\hline
				\end{tabular}
		}}
		
	\end{center}
	\vspace{-0.2cm}
	\caption{$2A$ orbifold, non-standard embedding, with single Wilson line. Shift is given by, $\gamma=\{3,1,1,1,1,1,0^2\},\tilde{\gamma}=\{2,0^7\}$ and $\gamma=\{3,1,1,1,1,1,0^2\},\tilde{\gamma}=\{2,2,2,0^5\}$  }
	\renewcommand{\arraystretch}{0.5}
\end{table}

\begin{table}[H]
	\renewcommand{\arraystretch}{0.5}
	\begin{center}
		\vspace{0.5cm}
		\setlength{\doublerulesep}{20\arrayrulewidth}
		{\scriptsize{
				\begin{tabular}{|c|c|c|c|c|c|}
					\hline
					& & & & & \\
					$(n_1,n_2,b)$ & $(0,0,0)$& $(1,1,0)$ & $(2,1,0)$ & $(1,2,0)$ & $(2,2,0)$  \\
					\hline
					& & & & & \\
					$n^0_{(n_1,n_2,b)}$  & 196 & 89248 & 4636736 & 4603136 & 2013905048   \\
					\hline
					\hline
					& & & & & \\
					$(n_1,n_2,b)$ & $(1/2,0,0)$ & $(1/2,2,0)$ & $(1/2,-1,0)$ & $(1/2,1,0)$ & $(3/2,1,0)$\\
					\hline
					& & & & & \\
					$n^0_{(n_1,n_2,b)}$ &  224 & 88384 & 4 & 6952 & 714456   \\
					\hline
					\hline
					& & & & & \\
					$(n_1,n_2,b)$ & $(0,0,1)$& $(1,1,1)$ & $(2,1,1)$ & $(1,2,1)$ & $(2,2,1)$  \\
					\hline
					& & & & & \\
					$n^0_{(n_1,n_2,b)}$ & $-16$ & 26944 & 1875032 & 1858960 &   1023115248 \\
					%			& & &   \\
					\hline
					\hline
					& & & & & \\
					$(n_1,n_2,b)$ &  $(1/2,0,1)$ & $(1/2,2,1)$ & $(1/2,-1,1)$ & $(1/2,1,1)$ & $(3/2,1,1)$   \\
					\hline
					& & & & & \\
					$n^0_{(n_1,n_2,b)}$ & 48 & 27520 & 0 & 1376 & 257408   \\
					\hline
				\end{tabular}
		}}
		
	\end{center}
	\vspace{-0.2cm}
	\caption{$2A$ orbifold, non-standard embedding, with single Wilson line. 
		Shift is given by, $\gamma=1,1,0^6, \tilde{\gamma}=\{1,1,1,1,1,1,1,-1\}$ }
	\renewcommand{\arraystretch}{0.5}
\end{table}

\begin{table}[H]
	\renewcommand{\arraystretch}{0.5}
	\begin{center}
		\vspace{0.5cm}
		\setlength{\doublerulesep}{20\arrayrulewidth}
		{\scriptsize{
				\begin{tabular}{|c|c|c|c|c|c|}
					\hline
					& & & & & \\
					$(n_1,n_2,b)$ & $(0,0,0)$& $(1,1,0)$ & $(2,1,0)$ & $(1,2,0)$ & $(2,2,0)$  \\
					\hline
					& & & & & \\
					$n^0_{(n_1,n_2,b)}$  & 228 & 87456 & 4618816 & 4618368 & 2014718808   \\
					\hline
					\hline
					& & & & & \\
					$(n_1,n_2,b)$ & $(1/2,0,0)$ & $(1/2,2,0)$ & $(1/2,-1,0)$ & $(1/2,1,0)$ & $(3/2,1,0)$\\
					\hline
					& & & & & \\
					$n^0_{(n_1,n_2,b)}$ & 240 & 87456 & 0 & 7168 & 718848   \\
					\hline
					\hline
					& & & & & \\
					$(n_1,n_2,b)$ & $(0,0,1)$& $(1,1,1)$ & $(2,1,1)$ & $(1,2,1)$ & $(2,2,1)$  \\
					\hline
					& & & & & \\
					$n^0_{(n_1,n_2,b)}$ &  0 & 26688  & 1868088 & 1867776 & 1023639552 \\
					%			& & &   \\
					\hline
					\hline
					& & & & & \\
					$(n_1,n_2,b)$ &  $(1/2,0,1)$ & $(1/2,2,1)$ & $(1/2,-1,1)$ & $(1/2,1,1)$ & $(3/2,1,1)$   \\
					\hline
					& & & & & \\
					$n^0_{(n_1,n_2,b)}$ & 8 & 26688 & 0 & 1536  & 260096    \\
					\hline
				\end{tabular}
		}}
		
	\end{center}
	\vspace{-0.2cm}
	\caption{$2A$ orbifold, non-standard embedding, with single Wilson line. Shift is given by, $\gamma=\{3,1,0^6\}, \tilde{\gamma}=\{1,1,1,1,1,1,1,-1\}$ }
	\renewcommand{\arraystretch}{0.5}
\end{table}

\subsection{Non-standard embeddings of orbifold of order 3}

\begin{table}[H]
	\renewcommand{\arraystretch}{0.5}
	\begin{center}
		\vspace{0.5cm}
		\setlength{\doublerulesep}{20\arrayrulewidth}
		{\scriptsize{
				\begin{tabular}{|c|c|c|c|c|c|}
					
					\hline
					& & & & & \\
					$(n_1,n_2,b)$ & $(0,0,0)$ & $(1,1,0)$ & $(1,2,0)$ & $(1,3,0)$ & $(1,4,0)$ \\
					\hline
					& & & & & \\
					$n^0_{(n_1,n_2,b)}$ & $-24$ & 58374 & 3080196 &  79215060 & 1343181204  \\
					\hline
					\hline
					& & & & & \\
					$(n_1,n_2,b)$ & $(r/3,0,0)$ & (1/3,3,0) & (2/3,3,0) & (1/3,9,0) & (2/3,6,0) \\
					\hline
					& & & & & \\
					$n^0_{(n_1,n_2,b)}$ & 144 & 58374 & 3080196  & 79256142 & 1343181204 \\
					\hline
					\hline
					& & & & & \\
					$(n_1,n_2,b)$ & $(1/3,-1,0)$ & $(2/3,-1,0)$ & $(1/3,1,0)$ & (2/3,1,0) & (4/3,1,0)  \\
					\hline
					& & & & & \\
					$n^0_{(n_1,n_2,b)}$ & 0 & $18$ & 1674 & 11916 & 244188  \\
					\hline
					& & & & & \\
					$(n_1,n_2,b)$ & $(0,0,1)$ & $(1,1,1)$ & $(1,2,1)$ & $(1,3,1)$ & $(1,4,1)$ \\
					\hline
					& & & & & \\
					$n^0_{(n_1,n_2,b)}$ & $32$ & 18216 & 1247508 & 3685782 & 682471332   \\
					\hline
					\hline
					& & & & & \\
					$(n_1,n_2,b)$ & $(r/3,0,1)$ & (1/3,3,1) & (2/3,3,1) & (1/3,9,1) & (2/3,6,1) \\
					\hline
					& & & & & \\
					$n^0_{(n_1,n_2,b)}$ & 36 & 18216  & 1247508  & 36875376 & 682471332  \\
					\hline
					\hline
					& & & & & \\
					$(n_1,n_2,b)$ & $(1/3,-1,1)$ & $(2/3,-1,1)$ & $(1/3,1,1)$ & (2/3,1,1) & (4/3,1,1)  \\
					\hline
					& & & & & \\
					$n^0_{(n_1,n_2,b)}$ & 0 & $0$ & 216 & 2952 & 83592  \\
					\hline
				\end{tabular}
		}}
	\end{center}
	\vspace{-0.5cm}
	\caption{$3A$ orbifold, non-standard embedding with single Wilson line. Shift is given by, $\gamma=\{1,-1,0^6\}, \tilde{\gamma}=\{2,1,1,0^5\}$}
	\renewcommand{\arraystretch}{0.5}
\end{table}

\begin{table}[H]
	\renewcommand{\arraystretch}{0.5}
	\begin{center}
		\vspace{0.5cm}
		\setlength{\doublerulesep}{20\arrayrulewidth}
		{\scriptsize{
				\begin{tabular}{|c|c|c|c|c|c|}
					
					\hline
					& & & & & \\
					$(n_1,n_2,b)$ & $(0,0,0)$ & $(1,1,0)$ & $(1,2,0)$ & $(1,3,0)$ & $(1,4,0)$ \\
					\hline
					& & & & & \\
					$n^0_{(n_1,n_2,b)}$ & $60$ & 59778 & 3088044 &  79190148 & 1343316636  \\
					\hline
					\hline
					& & & & & \\
					$(n_1,n_2,b)$ & $(r/3,0,0)$ & (1/3,3,0) & (2/3,3,0) & (1/3,9,0) & (2/3,6,0) \\
					\hline
					& & & & & \\
					$n^0_{(n_1,n_2,b)}$ & 324 & 59778 & 3088044  & 79291062 & 1343181204 \\
					\hline
					\hline
					& & & & & \\
					$(n_1,n_2,b)$ & $(1/3,-1,0)$ & $(2/3,-1,0)$ & $(1/3,1,0)$ & (2/3,1,0) & (4/3,1,0)  \\
					\hline
					& & & & & \\
					$n^0_{(n_1,n_2,b)}$ & 54 & $0$ & 1458 & 12420 & 242028  \\
					\hline
					& & & & & \\
					$(n_1,n_2,b)$ & $(0,0,1)$ & $(1,1,1)$ & $(1,2,1)$ & $(1,3,1)$ & $(1,4,1)$ \\
					\hline
					& & & & & \\
					$n^0_{(n_1,n_2,b)}$ & 112 &  18144 & 1248048 & 36812608 & 682507296  \\
					\hline
					\hline
					& & & & & \\
					$(n_1,n_2,b)$ & $(r/3,0,1)$ & (1/3,3,1) & (2/3,3,1) & (1/3,9,1) & (2/3,6,1) \\
					\hline
					& & & & & \\
					$n^0_{(n_1,n_2,b)}$ & 0 & 18144 & 1248048 & 36881568 & 682507296  \\
					\hline
					\hline
					& & & & & \\
					$(n_1,n_2,b)$ & $(1/3,-1,1)$ & $(2/3,-1,1)$ & $(1/3,1,1)$ & (2/3,1,1) & (4/3,1,1)  \\
					\hline
					& & & & & \\
					$n^0_{(n_1,n_2,b)}$ & 0 & 0 & 0 & 3024 & 81648  \\
					\hline
				\end{tabular}
		}}
	\end{center}
	\vspace{-0.5cm}
	\caption{$3A$ orbifold, non-standard embedding with single Wilson line. Shift is given by, $\gamma=\{2,1,1,1,1,0^3\}, \tilde{\gamma}=\{0^8\}$}
	\renewcommand{\arraystretch}{0.5}
\end{table}

\begin{table}[H]
	\renewcommand{\arraystretch}{0.5}
	\begin{center}
		\vspace{0.5cm}
		\setlength{\doublerulesep}{20\arrayrulewidth}
		{\scriptsize{
				\begin{tabular}{|c|c|c|c|c|c|}
					
					\hline
					& & & & & \\
					$(n_1,n_2,b)$ & $(0,0,0)$ & $(1,1,0)$ & $(1,2,0)$ & $(1,3,0)$ & $(1,4,0)$ \\
					\hline
					& & & & & \\
					$n^0_{(n_1,n_2,b)}$ & 108 & 58338  & 3079596  &  79238340  & 1343160684  \\
					\hline
					\hline
					& & & & & \\
					$(n_1,n_2,b)$ & $(r/3,0,0)$ & (1/3,3,0) & (2/3,3,0) & (1/3,9,0) & (2/3,6,0) \\
					\hline
					& & & & & \\
					$n^0_{(n_1,n_2,b)}$ & 156 & 58338 & 3079596  & 79251990 & 1343160684 \\
					\hline
					\hline
					& & & & & \\
					$(n_1,n_2,b)$ & $(1/3,-1,0)$ & $(2/3,-1,0)$ & $(1/3,1,0)$ & (2/3,1,0) & (4/3,1,0)  \\
					\hline
					& & & & & \\
					$n^0_{(n_1,n_2,b)}$ & 6 & 0 & 1746 & 11748 & 244908  \\
					\hline
					& & & & & \\
					$(n_1,n_2,b)$ & $(0,0,1)$ & $(1,1,1)$ & $(1,2,1)$ & $(1,3,1)$ & $(1,4,1)$ \\
					\hline
					& & & & & \\
					$n^0_{(n_1,n_2,b)}$ & 8 & 17952 & 1245924 & 36861952 & 682440876   \\
					\hline
					\hline
					& & & & & \\
					$(n_1,n_2,b)$ & $(r/3,0,1)$ & (1/3,3,1) & (2/3,3,1) & (1/3,9,1) & (2/3,6,1) \\
					\hline
					& & & & & \\
					$n^0_{(n_1,n_2,b)}$ & 36 & 17952  & 1245924  & 36867840  & 682440876  \\
					\hline
					\hline
					& & & & & \\
					$(n_1,n_2,b)$ & $(1/3,-1,1)$ & $(2/3,-1,1)$ & $(1/3,1,1)$ & (2/3,1,1) & (4/3,1,1)  \\
					\hline
					& & & & & \\
					$n^0_{(n_1,n_2,b)}$ & $0$ & $0$ & 288 & 2928 & 84240  \\
					\hline
				\end{tabular}
		}}
	\end{center}
	\vspace{-0.5cm}
	\caption{$3A$ orbifold, non-standard embedding with single Wilson line. Shift is given by, $\gamma=\{2,0^7\}, \tilde{\gamma}=\{2,0^7\}$}
	\renewcommand{\arraystretch}{0.5}
\end{table}

\begin{table}[H]
	\renewcommand{\arraystretch}{0.5}
	\begin{center}
		\vspace{0.5cm}
		\setlength{\doublerulesep}{20\arrayrulewidth}
		{\scriptsize{
				\begin{tabular}{|c|c|c|c|c|c|}
					
					\hline
					& & & & & \\
					$(n_1,n_2,b)$ & $(0,0,0)$ & $(1,1,0)$ & $(1,2,0)$ & $(1,3,0)$ & $(1,4,0)$ \\
					\hline
					& & & & & \\
					$n^0_{(n_1,n_2,b)}$ & 96 & 58104  & 3078288  &  79242492  & 1343138112  \\
					\hline
					\hline
					& & & & & \\
					$(n_1,n_2,b)$ & $(r/3,0,0)$ & (1/3,3,0) & (2/3,3,0) & (1/3,9,0) & (2/3,6,0) \\
					\hline
					& & & & & \\
					$n^0_{(n_1,n_2,b)}$ & 126 & 58104 & 3078288  & 79246170 & 1343138112 \\
					\hline
					\hline
					& & & & & \\
					$(n_1,n_2,b)$ & $(1/3,-1,0)$ & $(2/3,-1,0)$ & $(1/3,1,0)$ & (2/3,1,0) & (4/3,1,0)  \\
					\hline
					& & & & & \\
					$n^0_{(n_1,n_2,b)}$ & 0 & 0 & 1782 & 11664 & 245268  \\
					\hline
					& & & & & \\
					$(n_1,n_2,b)$ & $(0,0,1)$ & $(1,1,1)$ & $(1,2,1)$ & $(1,3,1)$ & $(1,4,1)$ \\
					\hline
					& & & & & \\
					$n^0_{(n_1,n_2,b)}$ & 32 & 17964 & 1245834 & 36869488 & 682434882   \\
					\hline
					\hline
					& & & & & \\
					$(n_1,n_2,b)$ & $(r/3,0,1)$ & (1/3,3,1) & (2/3,3,1) & (1/3,9,1) & (2/3,6,1) \\
					\hline
					& & & & & \\
					$n^0_{(n_1,n_2,b)}$ & 18 & 17964  & 1245834  & 36866808  & 682434882  \\
					\hline
					\hline
					& & & & & \\
					$(n_1,n_2,b)$ & $(1/3,-1,1)$ & $(2/3,-1,1)$ & $(1/3,1,1)$ & (2/3,1,1) & (4/3,1,1)  \\
					\hline
					& & & & & \\
					$n^0_{(n_1,n_2,b)}$ & $0$ & $0$ & 342 & 2916 & 84564  \\
					\hline
				\end{tabular}
		}}
	\end{center}
	\vspace{-0.5cm}
	\caption{$3A$ orbifold, non-standard embedding with single Wilson line. Shift is given by, $\gamma=\{2,1,1,0^5\}, \tilde{\gamma}=\{2,1,1,1,1,0^3\}$}
	\renewcommand{\arraystretch}{0.5}
\end{table}

\section{Low lying coefficients}\label{lowcexp}
\subsection{${\cal Z}_{\rm new}$}\label{lowcf}

We list the low lying coefficients of the new supersymmetric index in this section for standard embeddings. We first note that for every orbifold of $K3$ studied here,
\bea
\sum_{s=0}^{N-1}c^{(0,s)}(-b^2/4)=2=\sum_{s=0}^{N-1}\tilde c^{(0,s)}(-b^2/4), \quad {\rm for} \; b^2=4\\ \nn
\sum_{s=0}^{N-1}c^{(0,s)}(-b^2/4)=112=\sum_{s=0}^{N-1}\tilde c^{(0,s)}(-b^2/4), \quad {\rm for} \; b^2=1
\eea
In each individual sector these are given by,
\bea
c^{(0,s)}(-b^2/4)=2/N=\tilde c^{(0,s)}(-b^2/4), \quad {\rm for} \; b^2=4\\ \nn
c^{(0,s)}(-b^2/4)=112/N=\tilde c^{(0,s)}(-b^2/4), \quad {\rm for} \; b^2=1
\eea
We also note that the Euler character is given by
\be
 \chi=\sum_b \sum_{s=0}^{N-1}c^{(0,s)}(-b^2/4)
 \ee
  for all the models studied in this paper.

\begin{table}[H]
	\renewcommand{\arraystretch}{0.5}
	\begin{center}
		\vspace{0.5cm}
		\begin{tabular}{|c|c|c|}
			\hline
			& & \\
			Orbifold & $c^{(0,1)}(0)$ & $\tilde c^{(0,1)}(0)$ \\
			\hline
			\hline
			& &  \\
			2A & $158$ & 136  \\
			3A & $142$ & 126 \\
			4B & 135 & 123\\
			5A & $542/5$ & 494/5  \\
			6A & 329/3 & 305/3\\
			7A & $601/7$ & 79\\
			8A & 165/2 & 153/2\\
			11A & $-656/11$ & 608/11\\
			14A/B & 715/14 & 667/14\\
			15A/B & 716/15 & 668/15\\
			23A & $718/23$ & 670/23 \\
			\hline
		\end{tabular}
	\end{center}
	\vspace{-0.2cm}
	\caption{$c^{(0,1)}(0),\; \tilde c^{(0,1)}(0)$ are listed  for different $N$.} \label{lowcoef1}
	\renewcommand{\arraystretch}{0.5}
\end{table}

Since the twisted elliptic genus for the orbifolds of composite order $N=mn$ in the $(0,s)$ sector where $s$ is composite can be related to the divisors of $N$ as in \cite{Chattopadhyaya:2017ews} hence it is sufficient for our purpose to list the $c^{(0,1)}(0)$ from the new-supersymmetric index to compute the perturbative polynomial. 

\subsection{$q$-expansion of Fourier transform of $F^{(r,s)}$}
\begin{eqnarray*}
\sum_{s=0}^1 F_{2A} ^{(0,s)}(\tau,z) &=& 2 \left(z+\frac{1}{z}+6\right)+\frac{4 q (z-3) (3 z-1) (z-1)^2}{z^2}+O(q^2)\\
 \sum_{s=0}^1 e^{-\pi is} F_{2A} ^{(0,s)}(\tau,z)&=& 8+ \frac{8 q ((z-6) z+1) (z-1)^2}{z^2}+O(q^2)=\sum_{s=0}^1 F_{2A} ^{(1,s)}(\tau,z)\\
\sum_{s=0}^1 e^{-\pi is} F_{2A} ^{(1,s)}(\tau,z)&=&-\frac{16 \sqrt{q} (z-1)^2}{z} +\frac{32 q^{3/2} (z-1)^2 ((z-4) z+1)}{z^2}+O(q^{5/2})
\end{eqnarray*}

\begin{eqnarray*}
\sum_{s=0}^2F_{3A}^{(0,s)} &=& 2 \left(z+\frac{1}{z}+4\right)+   \frac{4 q (z-1)^2 (z (2 z-7)+2)}{z^2}+O(q^2)\\
\sum_{s=0}^2 e^{-2\pi i k s/3} F_{3A}^{(0,s)} &=& 6+\frac{6 q (z-1)^2 ((z-5) z+1)}{z^2}+O(q^2), \quad k=1,2\\
\sum_{s=0}^2 e^{-2\pi i s/3} F_{3A}^{(1,s)} &=& -\frac{6 {q}^{1/3} (z-1)^2}{z}+\frac{6 q^{4/3} (z-1)^2 (z (2 z-11)+2)}{z^2}+O(q^2)\\
\sum_{s=0}^2 e^{-2\pi i 2s/3} F_{3A}^{(1,s)} &=& -\frac{18 q^{2/3} (z-1)^2}{z}+\frac{36 (-1 + z)^2 (1 + (-3 + z) z) q^{5/3}}{z^2}+O(q^{2})
\end{eqnarray*}
Various relations exist among different $F^{(r,s)}$ and also their Fourier transforms.  Detailed list of the $F^{(r,s)}$ for $g'\in[M_{23}]$ orbifolds of $K3$ can be found in \cite{Chattopadhyaya:2017ews}.

\providecommand{\href}[2]{#2}\begingroup\raggedright\endgroup

%\bibliography{ref-col1}
%\bibliographystyle{JHEP} 

\end{document}